\documentclass{aa}
\usepackage{graphicx}

\usepackage{txfonts}
\usepackage{hyperref}
\hypersetup{
     colorlinks   = true,
     citecolor    = blue
}

\usepackage[symbol]{footmisc}

\usepackage{mathtools}

\newcommand{\kms}{\,km\,s$^{-1}$} 

\newcommand{\MLunit}{\,M$_\odot$\,/\,L$_\odot$}
\newcommand{\Hi}{\ion{H}{i}}
\newcommand{\MLb}{$ M/L_{\rm Bulge}$}
\newcommand{\MLd}{$ M/L_{\rm Disk}$}

\newcommand{\oiii}{[\ion{O}{iii}]$_{5007}$}

\begin{document} 
\def \Ha{H$\alpha$}
\def \Hb{H$\beta$}

\def \nii{[\ion{N}{ii}]}
\def \niia{[\ion{N}{ii}]$_{6548}$}
\def \niib{[\ion{N}{ii}]$_{6583}$}

\def \oii{[\ion{O}{ii}]}
\def \oiia{[\ion{O}{ii}]$_{3727}$}
\def \oiib{[\ion{O}{ii}]$_{3729}$}

\def \sfr{$\Sigma_{\mathrm{SFR}}$}

   \title{First spectroscopic study of ionized gas emission lines in the extreme low surface brightness galaxy Malin 1}

   \subtitle{}

   \author{Junais
            \inst{1}, 
            S. Boissier\inst{1}, 
            B. Epinat\inst{1},
            P. Amram\inst{1},
            B. F. Madore\inst{2},
            A. Boselli\inst{1},
            J.Koda\inst{3},\\
            A. Gil de Paz\inst{4},
            J. C. Muños Mateos\inst{5},
            L. Chemin\inst{6}
            }

    \authorrunning{Junais}

  \institute{Aix Marseille Univ, CNRS, CNES, LAM, Marseille, France\\   
             \email{junais.madathodika@lam.fr}
             \and
             Observatories of the Carnegie Institution for Science, 813 Santa Barbara Street, Pasadena, CA 91101, USA
             \and
             Department of Physics and Astronomy, Stony Brook University, Stony Brook, NY 11794-3800, USA
             \and
             Departamento de Astrofísica, Universidad Complutense de Madrid, 28040 Madrid, Spain
             \and
             European Southern Observatory, Alonso de Cordova 3107, Vitacura, Casilla 19001, Santiago, Chile
             \and
             Centro de Astronomía (CITEVA), Universidad de Antofagasta, Avenida Angamos 601 Antofagasta, Chile
             }

   \date{Received December 19, 2019; accepted March 20, 2020}

 
  \abstract
  {Malin 1 is the largest known low surface brightness (LSB) galaxy, the archetype of so-called giant LSBs. The structure and the origin of such galaxies are still poorly understood, especially due to the lack of high-resolution kinematics and spectroscopic data.}
  {We use emission lines from spectroscopic observations of Malin 1
  aiming to bring new constraints on the internal dynamics and star formation history  of Malin 1.}
  {We have extracted a total of 16 spectra from different regions of Malin 1 and calculated the rotational velocities of these regions from the wavelength shifts and star formation rates from the observed \Ha{} emission line fluxes. We compare our data with existing data and models for Malin 1.}
  {For the first time we present the inner rotation curve of  Malin 1, characterized in the radial range $r$ < 10 kpc by a steep rise in the rotational velocity up to at least $\sim$350 km s$^{-1}$ (with a large dispersion), which had not been observed previously. We use these data to study a suite of new mass models for Malin 1. We show that in the inner regions dynamics may be dominated by the stars (although none of our models can explain the highest velocities measured) but that at large radii a massive dark matter halo remains necessary.
  The \Ha{} fluxes derived star formation rates are consistent with an early-type disk for the inner region, and with the level found in extended UV galaxies for the outer parts of the giant disk of Malin 1. We also find signs of high metallicity but low dust content for the inner regions.}
   {}

   \keywords{  Galaxies: individual: Malin1 -  Galaxies: kinematics and dynamics - Galaxies : star formation }
    \maketitle
%
\begin{figure*}
\centering
\begin{minipage}[b]{.475\textwidth}
\includegraphics[width=\hsize]{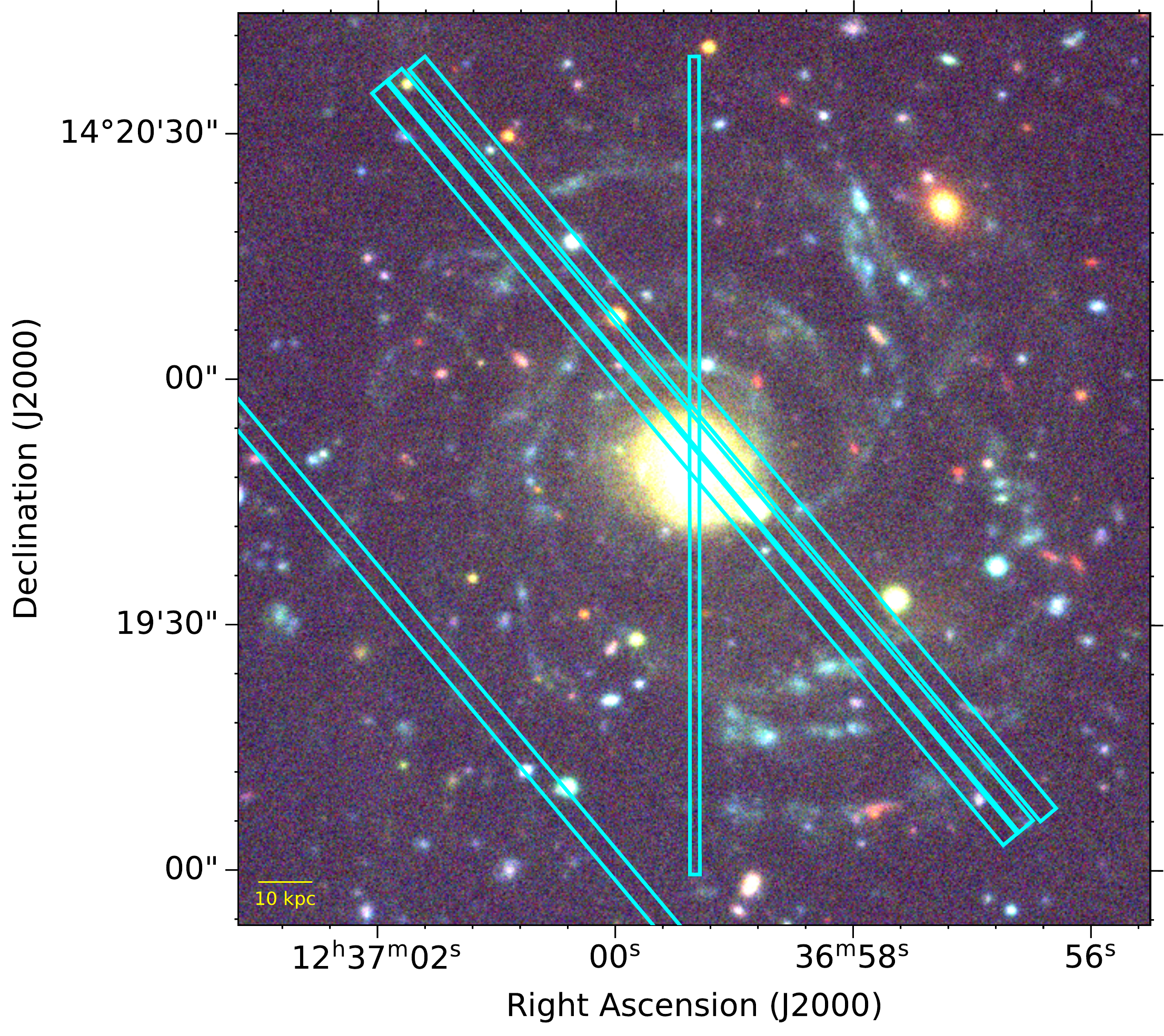}
\end{minipage}\qquad
\begin{minipage}[b]{.475\textwidth}
\includegraphics[width=\hsize]{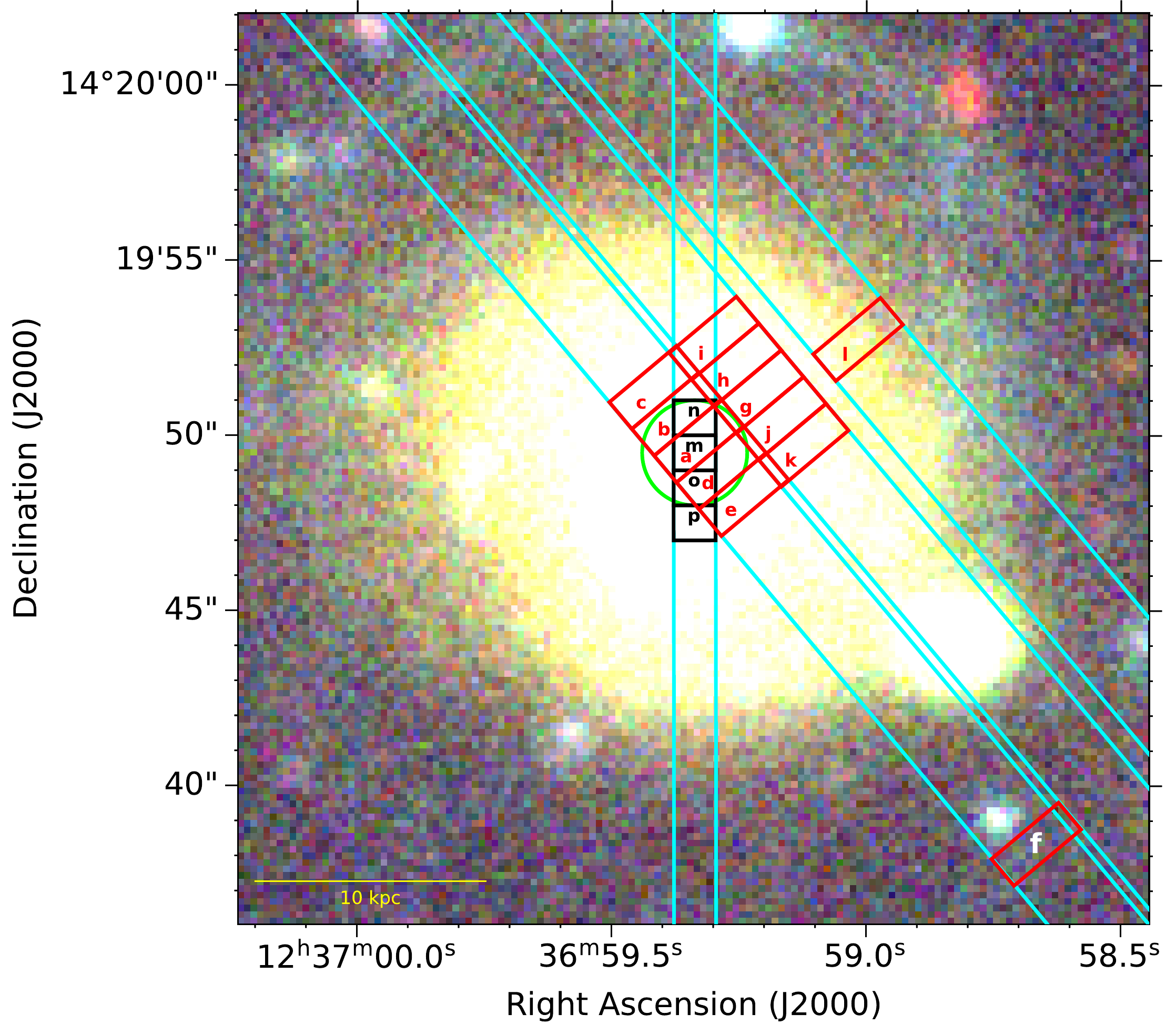}
\end{minipage}
\caption{\textit{Left}: Color composite image of Malin 1 from the CFHT-Megacam Next Generation Virgo cluster Survey (NGVS, \citealt{ferrarese}) \textit{u}, \textit{g} and \textit{i} band images. The slit positions of our observations are shown as blue rectangles. \textit{Right}:
Positions of the 16 apertures in which we could extract a spectrum. The 2016 and 2019 observations are marked as red and black regions respectively, along with their designated region names (see Tables. \ref{data} and \ref{data2}). The green circular region marked in the center is the location of a SDSS spectrum of Malin 1 with an aperture of $3\arcsec$ diameter.}
\label{slits}
\end{figure*}

\section{Introduction}
     The faint and diffuse galaxies that emit much less light per unit area than normal galaxies are known as Low Surface Brightness (LSB) galaxies. Although there is no clear-cut convention for defining LSB galaxies, they are usually broadly defined as galaxies with a disk central surface brightness ($\mu_0$) much fainter than the typical \cite{freeman} value for disk galaxies ($\mu_{0,B} = 21.65 \pm 0.30\,\, \text{mag arcsec}^{-2}$).
     LSBs may account for a very large galaxy population and Dark Matter (DM) content \citep{LSB,Blanton2005,de_blok}.
     Therefore, understanding this type of galaxies and their rotation curves, provided we have good kinematics data, could bring out some new insights on our current galaxy formation and evolution scenarios.
    
    LSB galaxies span a wide range in sizes, masses and morphology, from the largest existing galaxies down to the more common dwarfs. Giant Low Surface Brightness galaxies (GLSBs) are a sub-population of LSB galaxies, with an extremely extended low surface brightness disk with scale lengths ranging from $\sim$10 kpc to $\sim$50 kpc \citep{malin1_discovery}, and rich in gas content ($M_{HI}\sim10^{10}M_{\odot}$; \citealt{mathews}). Despite their low central surface brightness they are sometimes as massive as many "regular" galaxies \cite[see Figure~3 of][]{sprayberry}. 
    The origin of giant LSBs has been much debated with many propositions, e.g., face-on collisions \citep{mapelli2008}, cooling gas during a merger \citep{zhu,saburova2018}, large initial angular momentum \citep{boissier2003,Amorisco2016}, accretion from cosmic filaments \citep{saburova2019}. Few spectroscopic studies were possible in giant LSBs \citep[e.g.,][]{saburova2019}, while they can bring important information to distinguish between these possibilities.
    
    Malin 1 was discovered in 1986 \citep{malin1_discovery} and is the archetype of GLSB galaxies with a radial extent of $\sim$120 kpc \citep{moore_parker}. 
    The galaxy was accidentally discovered in the course of a systematic survey of the Virgo cluster region designed to detect extremely low surface brightness objects \citep{malin1_discovery}.
    It has an extrapolated disk central surface brightness of $\mu_{0,V}\approx25.5 \,\,\text{mag arcsec}^{-2} $ \citep{LSB}. However, despite its faint surface brightness disk, Malin 1 is a massive galaxy with a total absolute magnitude of $M_V \approx -22.9 \,\,\text{mag}$ \citep{pickering}. It is among the most gas rich galaxies known with an \Hi\ mass of $\sim 5 \times 10^{10} M_{\odot}$ \citep{pickering,mathews}.
    Malin 1 lies in a relatively low density environment in the large scale structure, typical for LSB galaxies \citep{reshetnikov}. Using the DisPerSE code \citep{Sousbie2011} with SDSS/BOSS data, we found Malin 1 lies at a distance of about 10 Mpc from the edge of its closest filament. This relatively low density but proximity to a filament could account for the stability and richness of its extremely huge gaseous disk.
    The analysis of a Hubble Space Telescope (HST) I-band image by \cite{barth} suggests that Malin 1 has a normal barred inner spiral disk embedded in a huge diffuse LSB envelope, making it similar to galaxies with an eXtended Ultraviolet (XUV) disk found in 30\% of nearby galaxies \citep{thilker}. Therefore Malin 1 can also be seen as the most extreme case of this class of galaxies. It is especially interesting to understand the nature of such disks when more extended galaxies have been found recently \citep{hagen,zhang18}.
    So far, limited spectroscopic data have been available for Malin 1. A full velocity map is provided by \citet{lelli}, but it is obtained from HI data, with a low spatial resolution. In the optical, a spectrum of the central 3 arcsec was obtained by SDSS (region shown in Fig. \ref{slits}) and is used by \citet{subramanian16} to analyze the Active Galactic Nuclei (AGN) properties of a sample of LSBs including Malin 1. Finally, \citet{reshetnikov} used obtained spectra along one long-slit passing by the center of Malin 1 and a companion, but they did not extracted from it an in-plane rotation curve of the galaxy, and rather concentrated on the possible interaction of Malin 1 and its companion.
    In this work we analyze new spectroscopic data concerning Malin 1. We derive new constraints on the inner kinematics and star formation rate surface densities within about 26 kpc. For the sake of comparison, we adopt the same cosmology as \citet{lelli} with $H_0 = 70 \,\,\text{km s}^{-1} \text{Mpc}^{-1}$,  $\Omega_{M} = 0.27$ and $\Omega_{\Lambda} = 0.73$, which corresponds to a projected angular scale of $1.56 \,\,\text{kpc arcsec}^{-1}$ and a distance of 377 Mpc. This cosmology is consistent with the ones found in modern cosmological experiments and close to the WMAP9 results \citep{Hinshaw2013}. 
    The basic properties of the galaxy adopted in this work are summarized in Table \ref{basic_properties}.
    In Sect. \ref{data_and_reduction} we discuss the data used in this work along with the steps followed for the data reduction. Section \ref{results} gives the major results we have obtained in this work. A detailed discussion on the consequences of our results along with a comparison of existing data and models is given in Sect. \ref{discussion}. Section \ref{mass_model} is dedicated to an extensive study of a suite of new Malin 1 mass models. Conclusions are given in Sect. \ref{conclusion}. 
    \begin{table}[h]
    \caption[]{\label{basic_properties}Selected properties of Malin 1}
    \begin{tabular}{lcc}
    \hline \hline
    Property &
    Value &
    References \\

    \\ \hline
    R.A. (J2000)   & 12$^\text{h}$ 36$^\text{m}$ 59.350$^\text{s}$ & 1\\
    Dec. (J2000)   & +14$^\circ$ 19$\arcmin$ 49.32$\arcsec$ & 1\\
    Redshift   & $0.0826 \pm 0.0017$   & 2\\
    V$_{\mathrm{sys}}$ (km s$^{-1}$)                  & $24766.7 \pm 4.0$      & 2 \\
    D$_{\text{L}}$ (Mpc)        & $377 \pm 8$     & 2\\
    Inclination Angle          & $38^\circ  \pm 3^\circ$ & 2\\
    Position Angle (PA)\footnotemark[2]  &      0$^\circ$   & 2\\
    
    \hline
    \noalign{\smallskip}
    
    \end{tabular}
    \tablebib{(1) NED Database; (2)~\citet{lelli}}
    \footnotemark[2]{\tiny{PA is adopted to be 0$\degr$ for the regions of Malin 1 within r < 26 kpc, where we have data in this work (see \citealt[Fig.~2]{lelli}})}
    \end{table}
%
   \begin{figure*}
   \resizebox{\hsize}{!}
            {\includegraphics{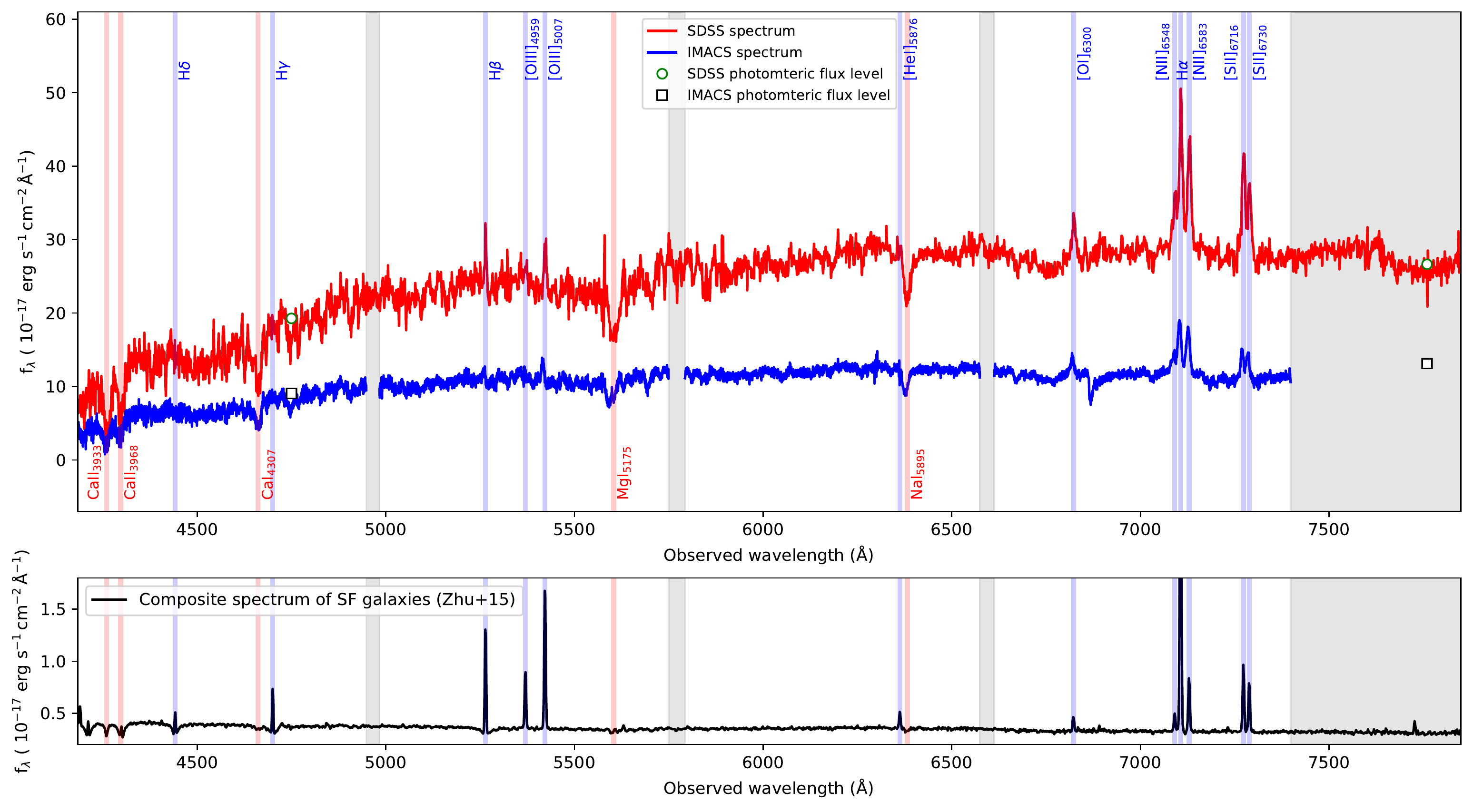}}
      \caption{\textit{Top:} Central region spectra of Malin 1. The blue curve is the spectrum extracted in  our central 1$\arcsec\times$ 2.5$\arcsec$ aperture (region \textit{a} in Fig. \ref{slits}). The red curve is the SDSS spectrum of the center of the galaxy, extracted within its circular optical fiber of diameter 3$\arcsec$ (shown in Fig. \ref{slits}). The green open circles and black open squares indicate the photometric flux levels obtained respectively within the SDSS and our aperture, using the NGVS \textit{g} and \textit{i} band photometric images of Malin 1.  The gray shaded area are the regions where we do not have data. \textit{Bottom:} For comparison, we show the median composite spectrum of all star forming galaxies (at $0 < z \lesssim 1.5$) from the SDSS eBOSS observations \citep{Zhu2015}, shifted to the redshift of Malin 1. The blue and red vertical shaded regions indicate the main identified emission and absorption lines. }
         \label{spectra}
   \end{figure*}    
%
%
    \begin{figure*}[ht!]
    \centering
    \begin{minipage}[b]{\textwidth}
    \includegraphics[width=\hsize]{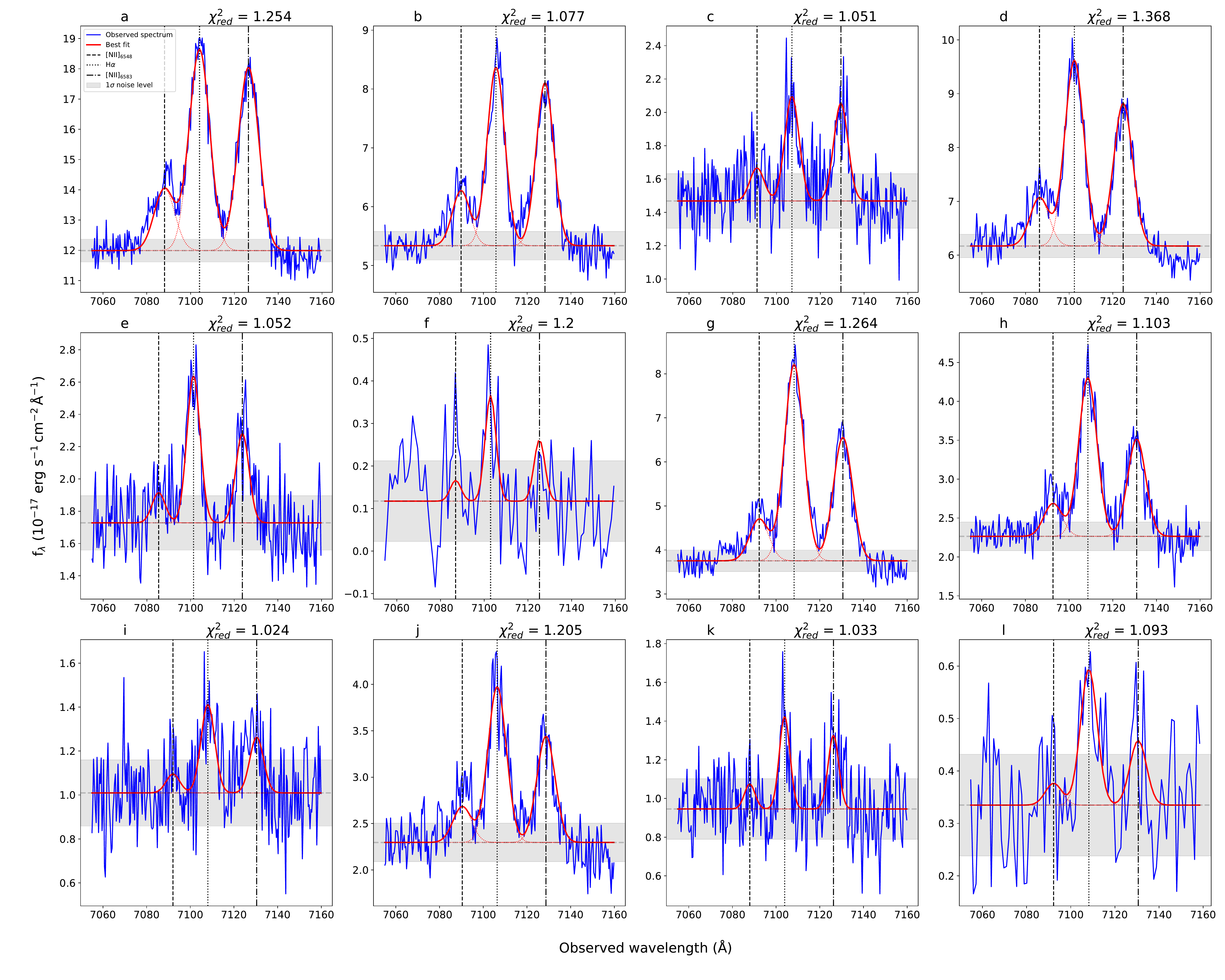}
    \end{minipage}\qquad
    \begin{minipage}[b]{\textwidth}
    \includegraphics[width=\hsize]{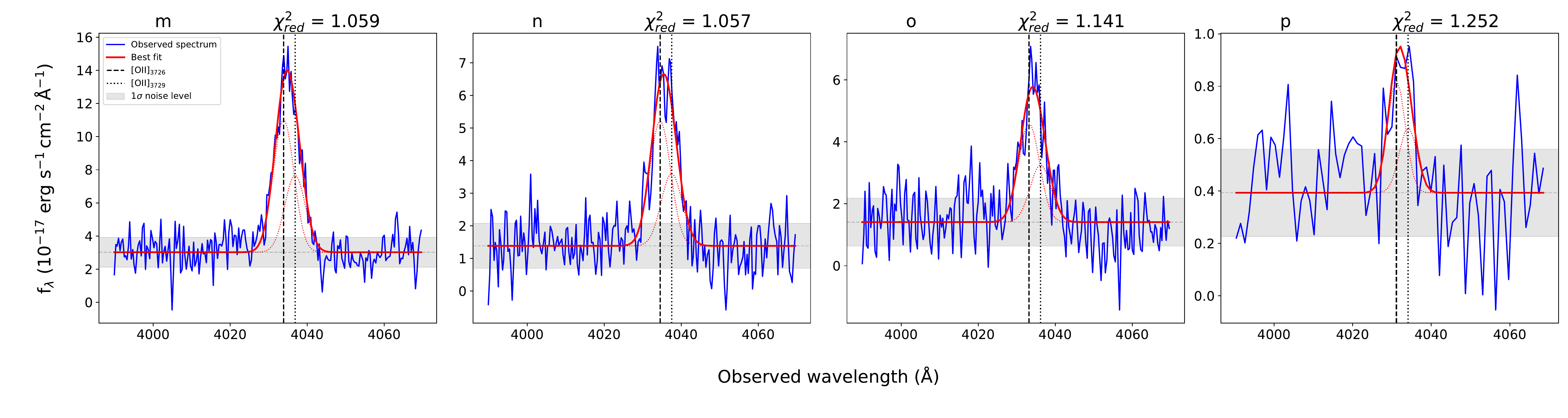}
    \end{minipage}
    \caption{Zoom on the wavelength range of interest for the 16 spectra extracted in this work (12 \Ha{} spectra in the top three rows and 4 \oii{} spectra in the fourth row). The solid red curve is the best fit along with its decomposition in single lines shown as thin red dotted lines. The gray dashed line and shaded region indicates the continuum level obtained from the fitting with the $1\sigma$ noise level. The black dashed, dotted and dot-dash vertical lines marks the positions of the \niia{}, \Ha{} and \niib{} emission lines respectively for the three top rows. The dashed and dotted lines in the bottom row show the position of the two components of the [OII] doublet. The region name and the reduced $\chi^{2}$ are indicated on top of the each panel.}  
    \label{fittting_results}
    \end{figure*}


    \begin{table*} 
    \scalebox{0.94}{
    \begin{tabular}{cccccccc} \hline \hline 
    & & \multicolumn{1}{c}{\underline{\hspace{0.6cm}Wavelength\hspace{0.6cm}}} &
    \multicolumn{1}{c}{\underline{\hspace{0.6cm}Velocity\vphantom{g}\hspace{0.6cm}}}&
    \multicolumn{4}{c}{\underline{\hspace{4.25cm}Flux\vphantom{g}\hspace{4.25cm}}}\\ 
    Region & Radius & $\lambda_{H\alpha}$ & V$_{\text{rot}}$ & \Ha{} & \niib{} & \Hb{} & \oiii\\
    
   name & (kpc) & (\AA) & (km s$^{-1}$) & \hspace{2cm} & \hspace{2cm} & \hspace{2cm} & \hspace{2cm} \\\\
    (1) & (2) & (3) & (4) & (5) & (6) & (7) & (8)\\ \hline \\
    a      &  0.58 $\pm$ 0.38    & 7104.07 $\pm$ 0.08    &   -53 $\pm$ 54       &  78.3 $\pm$ 1.5     &  71.4  $\pm$ 1.6    & 10.8 $\pm$ 1.2  & 25.7 $\pm$ 1.7\\
    
    b   &   2.00 $\pm$ 0.37  &  7105.77  $\pm$ 0.09  &     80 $\pm$ 23    &  30.5  $\pm$ 0.9     &  27.9 $\pm$  0.9    &   4.7 $\pm$ 0.7 & ...\\  
    
    c   &   3.70 $\pm$ 0.40  &  7107.04  $\pm$ 0.27  &     197 $\pm$ 39    &  5.2 $\pm$ 0.6       &  4.8 $\pm$ 0.6      &    ... & 1.7 $\pm$ 0.5 \\
    
    d   &   1.64 $\pm$ 0.48  &  7102.38 $\pm$  0.09  &     408 $\pm$ 173    &  37.8 $\pm$ 0.9      &  29.1 $\pm$ 0.9     &  7.0 $\pm$ 0.9 & 12.4 $\pm$ 0.9\\

    e   &  3.33 $\pm$ 0.46   &  7101.37 $\pm$ 0.22   &     420 $\pm$ 87    &  7.0 $\pm$ 0.6       &  4.2  $\pm$  0.5    & 3.5 $\pm$ 1.1 & 2.7 $\pm$ 0.6\\

    f$^\text{*}$   &  25.9 $\pm$ 0.43   &  7102.93 $\pm$ 0.88   &     189 $\pm$ 91    & 1.6 $\pm$  0.5       & 1.0 $\pm$ 0.5       &  ... & ... \\  

    \\\\ 
        
    g    & 4.53  $\pm$ 0.20   &  7108.22 $\pm$ 0.08   &     393 $\pm$ 24   &   48.9 $\pm$ 1.0     & 30.7   $\pm$ 0.9    & 11.2  $\pm$ 0.6 & 9.1 $\pm$ 0.7\\  
   
    h    & 4.56 $\pm$ 0.18    &  7108.62 $\pm$ 0.13   &     308 $\pm$ 13   &   21.2 $\pm$ 0.7     &  13.0  $\pm$ 0.7   & 3.1 $\pm$  0.6 & 3.5 $\pm$ 0.6\\ 

    i    & 5.21 $\pm$  0.14   &   7108.02 $\pm$ 0.47  &     228 $\pm$ 33   &   3.2 $\pm$ 0.5      &   2.0  $\pm$ 0.5    &  ... & 1.9 $\pm$ 0.5\\
        
    j    & 5.12 $\pm$ 0.19    &   7106.38  $\pm$ 0.17 &     367 $\pm$ 54   &   17.7 $\pm$ 0.8     &  12.1  $\pm$  0.8   &  3.5 $\pm$ 0.8 & 5.6 $\pm$ 0.7\\

    k    & 6.16 $\pm$ 0.17    &   7103.89 $\pm$ 0.33  &     -1092 $\pm$ 692   &   3.0 $\pm$  0.5    &  2.4 $\pm$  0.4     &   ... & ... \\  

    \\\\

   l$^\text{*}$    &  10.51 $\pm$  0.27  &  7108.20 $\pm$ 1.08  &     489 $\pm$ 155  &     2.5 $\pm$ 0.7   &  1.2 $\pm$  0.8     &  ...  & 1.7 $\pm$ 0.4\\

    \\ \hline \hline \\\end{tabular}}\normalsize
     \scriptsize \caption{Extracted data for Malin 1 from the 2016 observation. (1) Name of the spectral extraction region. The * symbol indicates spectra that were re-binned for the analysis (see Sect. \ref{data_and_reduction}). (2) Radius in the galaxy plane. (3) \Ha{} observed wavelength. (4) Rotational velocity in the plane of the galaxy. (5-8) Observed flux of \Ha{}, \niib{}, H$_\beta$ and \oiii \,emission lines respectively within the 1$\arcsec\times$ 2.5$\arcsec$ regions. The flux units are in 10$^{-17}$ erg s$^{-1}$ cm$^{-2}$. The errorbars shown in the table include
     the positioning error (column 2), the fitting errors (column 3 and 5 to 8) or both (column 4) (see Appendix \ref{appendix_error_estimation}).}{\label{data}}
    \end{table*}



    \begin{table} 
    \scalebox{1}{
    \begin{tabular}{cccc} \hline \hline 
    & & \multicolumn{1}{c}{\underline{\hspace{0.4cm}Wavelength\hspace{0.4cm}}} &
    \multicolumn{1}{c}{\underline{\hspace{0.4cm}Velocity\vphantom{g}\hspace{0.4cm}}}\\ 
    Region & Radius & $\lambda_{[\ion{O}{ii}]3727}$ & V$_{\text{rot}}$  \\
    
    name & (kpc) & (\AA) & (km s$^{-1}$)\\\\
    (1) & (2) & (3) & (4) \\ \hline \\
    m  & 0.00 $\pm$ 0.64  &  4033.85 $\pm$ 0.11   &  23 $\pm$ 13 \\
    
    n & 1.56 $\pm$ 0.13 &  4034.52 $\pm$ 0.18   &  104 $\pm$ 23 \\
    
    o & 1.56 $\pm$ 0.13 &  4033.20 $\pm$ 0.24   &  55 $\pm$ 29 \\
    
    p$^\text{*}$  & 3.11 $\pm$ 0.07 &  4031.14 $\pm$ 0.67   &  304 $\pm$ 81 \\  

    \\ \hline \hline \\\end{tabular}}
     \caption{Extracted data for Malin 1 from the 2019 observation (2019 data are not flux calibrated). (1) Name of the spectral extraction region. The * symbol indicates spectra that were re-binned for the analysis (see Sect. \ref{data_and_reduction}). (2) Radius in the galaxy plane. (3) \oiia{} observed wavelength. (4) Rotational velocity in the plane of the galaxy. The errorbars shown in the table includes the positioning error (column 2), the fitting error (column 3), or both (column 4).}{\label{data2}}
    \end{table}


\section{Data \& Reduction}\label{data_and_reduction}

    The spectroscopic data of Malin 1 used in this work were obtained with the IMACS spectrograph at the 6.5m Magellan Baade telescope in the Las Campanas Observatory, Chile. Two runs of observation took place, in 2016 and 2019, with long-slits of width 2.5$\arcsec$ and 1.2$\arcsec$ respectively.
    
    In 2016, four slit positions were observed. We extracted a total of 12 spectra from different regions of size $1\arcsec\times2.5\arcsec$ each, for the three slit positions for which it was possible to obtain a clear signal. This includes a region at $\sim$26 kpc which is relatively far from the center of Malin 1 (see \figurename{ \ref{slits}}, region \textit{f}). These observations cover a wavelength range of 4250-7380 \AA \, with a dispersion of 0.378 \AA\,/pixel and a spectral resolution $R$ $\sim$850. The large width of the slit was chosen to optimize the chance of detecting \ion{H}{ii} regions within the slit. The orientation of the slits was chosen on the basis of UV images from \citet{boissier08}. Each of the slit positions had an exposure time of 3$\times$1200 seconds, oriented at a position angle of 39.95$\degr$ with respect to the major axis of the galaxy (see Fig. \ref{slits}). The initial position passes through the galaxy center. For subsequent positions, the slit positions were shifted from each other by a distance of 2.5$\arcsec$ towards the west (except for the 4th slit position that was moved about 50\arcsec\,towards east in order to pass through distant UV blobs, but we could not detect anything at this position). In order to obtain a precise position for each observations, we simulated the expected continuum flux along the slit, based on an image of Malin 1 acquired during the night of the observations (see Appendix \ref{appendix_error_estimation}). A $\chi^{2}$ comparison with our spectral data allowed us to deduce the position of the slit. We estimated the position uncertainty following \citet{avni} and found it to be on the order of 0.1$\arcsec$ ($99\%$ confidence level).
    This process resulted in a small overlap for the slit positions 1 and 2, that is, however,
    negligible considering the size of the apertures in which we extracted our spectra and thereby each of our apertures are considered as independent regions.\\ 
    The 2019 observation of Malin 1 was performed using a narrower slit-width of 1.2$\arcsec$ oriented at a position angle of 0$\degr$, along the major axis of the galaxy. The observations were done for a single position with an exposure time of 2$\times$1200 seconds and a wavelength coverage of 3650-6770 \AA \, to obtain a spectral resolution R $\sim$1000. We extracted  4 spectra from this run for which the \oii{} doublet ($\lambda$3727,3729) is clearly detected (although the two lines overlap at our resolution), each with an aperture size of $1\arcsec\times1.2\arcsec$. The average atmospheric seeing measured at the location was $\sim$1$\arcsec$, with an airmass of 1.4 and a spatial sampling of 0.111$\arcsec$/pixel for all the Malin 1 observations.\\

    For both runs, the data reduction and the spectral extraction was carried out using standard IRAF\footnote{IRAF is distributed by the National Optical Astronomy Observatory, which is operated by the Association of Universities for Research in Astronomy (AURA) under a cooperative agreement with the National Science Foundation.}
    tasks within the \textit{ccdred} and \textit{onedspec} packages. The wavelength calibration was done using a standard HeNeAr arc lamp for each aperture independently. 
    Flux calibration of the extracted spectra from the 2016 observation was done using the reference star LTT 3218 (observed at an airmass of 1.011). We did not perform a flux calibration for the 2019 data, since we do not have a proper reference star for this observation. For illustration purpose, as shown in the fourth row of Fig. \ref{fittting_results}, we normalized the flux with the NGVS \textit{u}-band photometry in the same aperture as our slit. Since this is not a proper calibration,  we do not provide line fluxes in this case. 
    However, we checked that if we adjust the continuum level with this \textit{u}-band photometry (or with the overlapping spectra from 2016 observations), the \oiia{}/\Ha{} flux ratio obtained is within the values found by \citet{Mouhcine2005}. 
    
    We were able to extract a total of 16 spectra from different regions of Malin 1 (indicated in \figurename{ \ref{slits}}) where it was possible to obtain a clear signal for our target emission lines (\Ha{} and \oii{}). For each slit position, we started from the peak of emission and moved outwards until no signal was measured around the expected line position (the naming of each region shown in Fig. \ref{slits} is based on this). This allowed us to obtain 15 measurements in the central region. 
    We kept any regions for which the peak of the emission line is visible above the noise (above about $2\sigma$). When it is the case, we find we could fit a line (sometimes after spectral re-binning for few regions, as explained further).
    The inner part of the galaxy, as visible in the broad-band image (Fig. \ref{slits}), is more extended than the regions for which we could secure a detection. This is because the central region is basically similar to an early-type disk \citep{barth}  with old stars but little gas, thus the emission signal drops quickly to very low level.

    After extracting 15 spectra in the inner part of the galaxy, we inspected the rest of the galaxy where we had data and searched for emission, especially those regions close to spiral arms observed in optical wavelength or blobs in the GALEX UV images of Malin 1 \citep{boissier16}. For this, we used apertures of the same size as the one applied in the inner galaxy, but also larger apertures in order to increase the signal to noise ratio in case of extended emission. However, we were able to recover only one additional spectrum $\sim$26 kpc away from the center, close to a compact source visible in the broadband images from NGVS \citep{ferrarese}, as shown in \figurename{ \ref{slits}}. It also coincides with a UV blob from the GALEX images of Malin 1. We checked that some of the UV emission overlaps with our aperture, however, the GALEX resolution of about 5\arcsec make this association uncertain. We thus turned to UVIT \citep{Kumar2012} images of Malin 1 that became recently publicly available and we still found some UV emission at this position (at the UVIT resolution of 1.8$\arcsec$, close to the size of our aperture).

    We have focused on the \Ha{} and \oii{} emission lines which were the strongest among those in the observed wavelength range (\Ha{} for the 2016 observation and \oii{} for the 2019 ones). For simplicity, we indifferently refer to the 2016 and 2019 observations as to the \Ha{} and \oii{} ones.
    
    We performed a fit of the emission lines using Python routines implementing a  Markov Chain Monte Carlo (MCMC) method on a Gaussian line profile in order to obtain the peak wavelength, flux and the associated errorbars of each emission line (see Appendix \ref{appendix_error_estimation}). Overlapping lines (\Ha{} and \nii{}; and the \oii{} doublet) are fitted simultaneously. Various constraints were applied on the emission lines during the fitting procedure, including a fixed line ratio for the \nii{} and \oii{} doublets (\niib{}/\niia{} $ = 2.96$ adopted from \citealt{ludwig}; \oiib{}/\oiia{} $ = 0.58$ from \citealt{Pradhan2006, Comparat2016}).  
    The line ratio of the \oii{} doublet do depend on the electron density. We performed tests with values covering the range 0.35 to 1.5 and finally adopted the typical ratio of 0.58, since we do not know for sure the physical conditions in galaxies of very low surface brightness, and the choice was not affecting our conclusions. 
    We also fixed the line separations using the laboratory air wavelengths of the emission lines and taking into account the redshift (given in Table \ref{basic_properties}), 
    where $\Delta\lambda_{obs}$ = $\Delta\lambda_{lab}\mathrm(1+z)$. 
    The uncertainty in the redshift is negligible (within $1\sigma$ error of the wavelength and flux values) and does not affect our results. We also performed a spectral re-binning by a factor 3 for a few of our observations affected by a considerably weaker signal that are at the limit of our detection (regions \textit{f}, \textit{l} and \textit{p} from Fig. \ref{fittting_results}), in order to increase the signal to noise ratio (this allows us to secure a measurement in these apertures, at the price of a lower spectral resolution). The detailed results of our fitting procedure are shown in \figurename{ \ref{fittting_results}}, Table \ref{data} and Table \ref{data2}.  

    The robustness of the spectral extraction was checked by the comparison of our central region spectrum (region \textit{a}) with that of an SDSS spectrum (DR12) of Malin 1 (see \figurename{ \ref{spectra}}).
    The entire range of both spectra are consistent in terms of the line positions and features.
    The continuum flux levels in both spectra are also consistent with the expected photometric flux levels measured within their corresponding apertures (shown in Fig. \ref{slits}) using NGVS \textit{g} and \textit{i} band images of Malin 1. For this central region, we also performed an underlying stellar continuum fit using the pPXF (Penalized Pixel-Fitting) method by \citet{Cappellari2017}, and also tried to include a broad component to take into account he nucleus activity. However, the stellar continuum subtraction together with the additional active nucleus \Ha{} broad-line component modified our emission line measurement results by less than $1.5\sigma$.
    Since the continuum is too noisy in many of the apertures, it would not be possible to fit it with pPXF in each apertures. Since in the center where those effects should be the largest, they do not affect our conclusion, we choose to adopt the same procedure in each apertures, i.e. not fitting the underlying stellar continuum subtraction and broad line component in the results presented in the paper (Table \ref{data}).

\section{Results}\label{results}

    \subsection{Rotation curve}\label{rc_geometry}

        \begin{figure*}[ht!]
        \centering
        \begin{minipage}[b]{.475\textwidth}
        \includegraphics[width=\hsize]{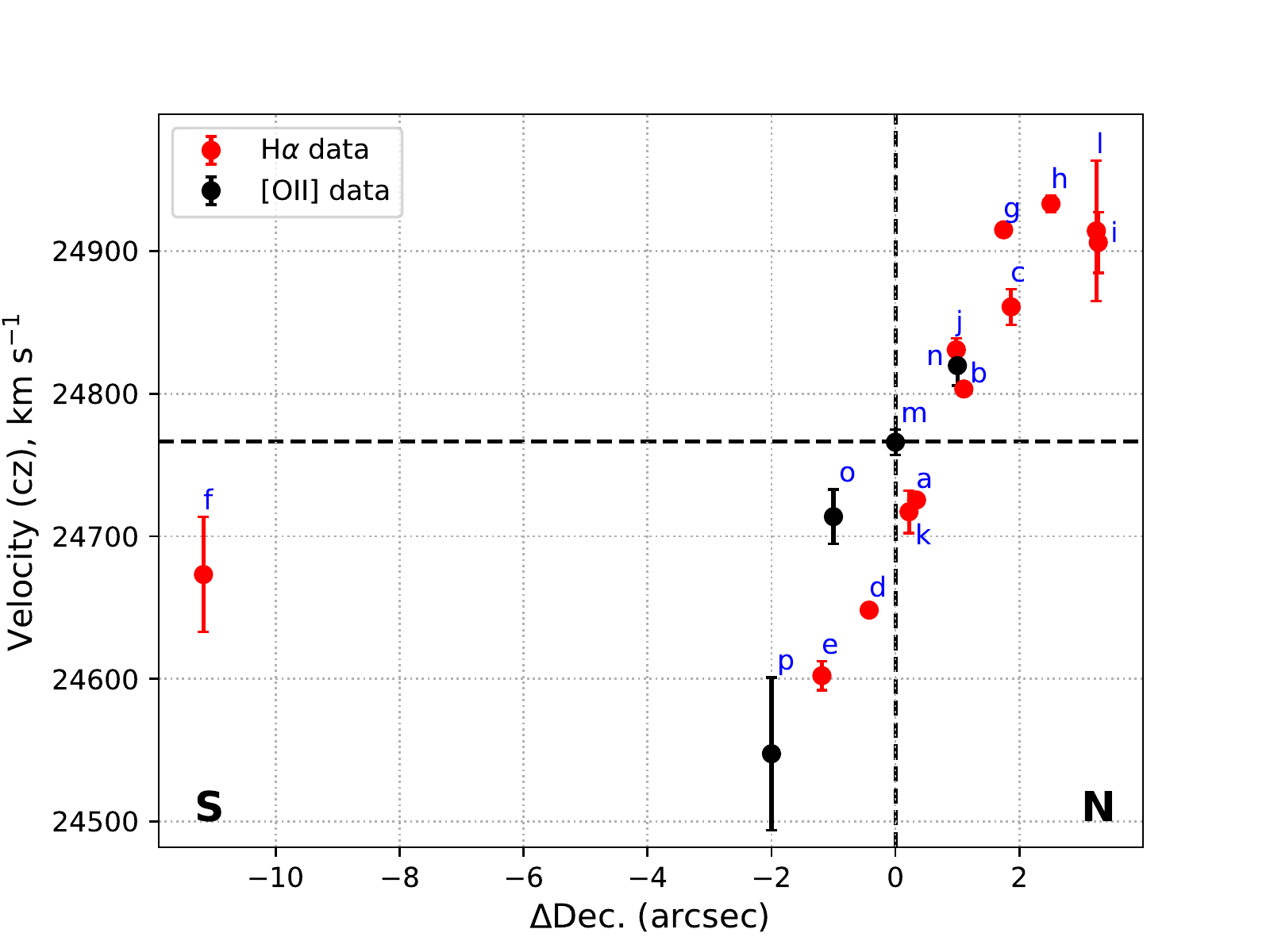}
        \end{minipage}\qquad
        \begin{minipage}[b]{.475\textwidth}
        \includegraphics[width=\hsize]{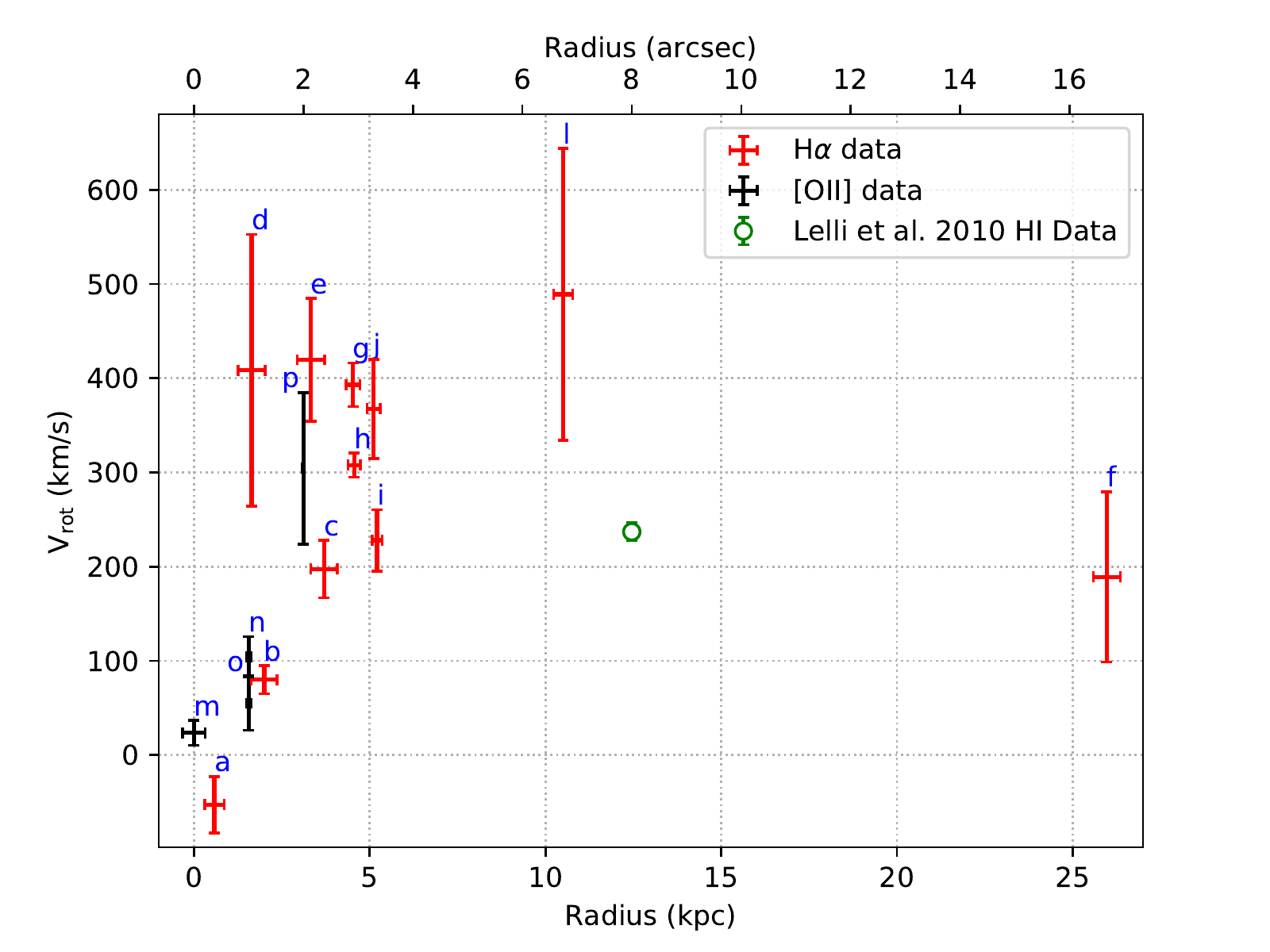}
        \end{minipage}
        \caption{\textit{Left}: Line of sight velocity measured from the observed shift in wavelengths for the \Ha{} and \oii{} lines (see Table \ref{data} and Table \ref{data2}). The x-axis corresponds to the projected radius on the major axis of the galaxy in the plane of sky. The thick horizontal and vertical dashed lines mark the V$_{\mathrm{sys}}$ and major axis of the galaxy respectively. \textit{Right}: Rotation curve of Malin 1, projected on the plane of the galaxy. The Red and black points indicate the \Ha{} data and \oii{} data respectively. Green open circle show the \citet{lelli} \Hi\ data point in the same radial range. The region name of each point is marked as blue letters.}
        \label{rc}
        \end{figure*}

        Rotation curves in LSB galaxies have long been debated \citep[see, e.g.,][]{de_blok,pickering,lelli}. The
        analysis of rotation curves is of utmost importance in understanding the dynamics and underlying mass distribution and may help to understand the origin of giant LSBs \citep{saburova2019}. One of the main results of this work is the extraction of a rotation curve for Malin 1 using the observed wavelength of \Ha{} and \oiia{} emission lines at different positions within the galaxy.

        The global systemic velocity (V$_{\mathrm{sys}}$) of Malin 1 was adopted from \citet{lelli} using \Hi\ measurements (see \tablename{ \ref{basic_properties}}) which is consistent with the velocity we measure in our Malin 1 center observation. The observed velocity shift at different regions of the galaxy from V$_{\mathrm{sys}}$ is used for the calculation of the rotational velocities on the galaxy plane as a function of radius. We applied a correction for the galaxy inclination angle and PA (see \tablename{ \ref{basic_properties}}), assuming an axi-symmetric geometry and a thin disk. However, a region too close to the minor axis of the galaxy (region \textit{k}) was eliminated from the rotation curve since it has a huge azimuthal correction ($\cos{\theta}$ = 0.05 $\pm$ 0.02) when re-projecting the observed velocity to the plane of the galaxy (see \tablename{ \ref{data}}). An additional correction for the heliocentric velocity due to the Earth's motion at the time and location of the observations were also added to the observed velocities (V$_{\mathrm{helio}}$ = -9.1 km s$^{-1}$ and -16.3 km s$^{-1}$ for the 2016 and 2019 data respectively). 
        
        The uncertainties on the velocities were computed by propagating the line of sight velocity measurements using the projection parameters of Malin 1 (line of sight and azimuthal deprojections).
        We quadratically added to this uncertainty the one related to the slit
        positioning described in Sect. \ref{data_and_reduction}. The impact of these uncertainties on the deprojected rotation velocities was computed using 10000 Monte Carlo realizations.
        The inclination and position angle that we adopted are also uncertain.
        However, we do not take in to account these uncertainties, since changing the inclination does not affect much the relative position of the points in the rotation curve. For example, a change of \textit{i} equal to $3\degr$ varies the rotation velocity varies by $\sim$15 km s$^{-1}$. A possible effect of uncertainty in inclination is discussed in \citet{lelli} as well. In addition, in this work, we combine the optical rotation curve with the \ion{H}{i} one from \citet{lelli}, so it is reasonable to use the same inclination and position angle as them.
        
        \figurename{ \ref{rc}} shows the extracted rotation curve of Malin 1. We observe a steep rise in the rotational velocity for the inner regions (inside $\sim$10 kpc) up to $\sim$350 km s$^{-1}$ (with, however, some spread between 200 and 400 km s$^{-1}$ around a radius of 5 kpc), and a subsequent decline to reach the plateau observed on large scales with \ion{H}{i}. 
        Such very high velocities (up to 570 km s$^{-1}$) are observed in massive spirals \citep{Ogle2019}.
        Both the \Ha{} and \oii{} velocities in our data appears to follow a similar trend and are consistent with each other. 
        A steep inner rise of rotation curve is typical for an High Surface Brightness (HSB) system. For Malin 1, it is the first time that we have observed this behavior, unlike the slowly rising rotation curve predicted by \cite{pickering} or the poorly resolved inner rotation curve from \cite{lelli} using \Hi\ data.
        
        The implications of this result and a comparison to existing models and data are discussed in Sect. \ref{discussion}, and motivate the computation of new mass models (Sect. \ref{mass_model}).
        
    \subsection{\Ha{} Surface Brightness \& Star Formation Rate}
        \label{subsecSFRmeasurement}   
        We have extracted the \Ha{} flux for the 12 regions of Malin 1 discussed in Sect. \ref{data_and_reduction}. The observed flux was corrected for inclination and also for the Milky Way foreground galactic extinction \citep{Schlegel1998} using the standard \cite{cardelli} dust extinction law.
        We expect a low dust attenuation within Malin 1 itself, since LSB galaxies in general host very small amounts of dust \citep{hinz,rahman}.
        With our data, we can probe the effect of dust attenuation on the Balmer ratio, compared to its theoretical value in the absence of dust. 
        Indeed, we measured both the \Ha{} and \Hb{} fluxes in 8 apertures. The Balmer ratio is however also affected by the underlying stellar absorption. Since our data lack spectral resolution to actually measure it, or signal in the continuum to fit the stellar populations in all of our regions, we apply standard equivalent width (EW) corrections.  
        A large diversity of stellar underlying absorption EW for \Ha{} and \Hb{} is found in the literature \citep{Moustakas2006,Moustakas2010,Boselli2013}.
        We first choose to apply the equivalent width corrections that were measured by \citet{Gavazzi2011} for 5000 galaxies (EW \Ha{}$_{\mathrm{abs}}$ = 1.3 \AA ) 
        and \citet{Moustakas2010} for the representative SINGS sample (EW \Hb{}$_{\mathrm{abs}}$ = 2.5 \AA ). The Balmer ratio of our 8 regions is then found within 3$\sigma$ of the theoretical value of 2.86 for Case B recombination \citep{osterbrock}. The ratio is especially sensitive to the correction to the weaker \Hb{} line. In order to check the effect of our choice, if we adopt instead another value among the literature: EW \Hb{}$_{\mathrm{abs}}$ = 5.21 \AA\, \citep{Boselli2013}, the Balmer ratio in the 8 regions is now systematically below the theoretical value of 2.86. The \Ha{}$_{\mathrm{abs}}$ and \Hb{}$_{\mathrm{abs}}$ EW values adopted above from the literature are also consistent with the EW values we obtained from our stellar continuum pPXF fitting of the region \textit{a} discussed in Sect. \ref{data_and_reduction}. 
        Although our EW correction procedure is uncertain, both choice of correction for the underlying stellar absorption lead to Balmer ratio consistent with the absence of dust attenuation. Moreover, Malin 1 is also undetected in far-infrared with \textit{Spitzer} and \textit{Herschel}, which also indicates low attenuation \citep{boissier16}. Therefore we can reasonably assume that the \Ha{} flux we measured in this work is only weakly affected from dust attenuation within Malin 1.

        \begin{figure}[h]
        \centering
        \includegraphics[width=\hsize]{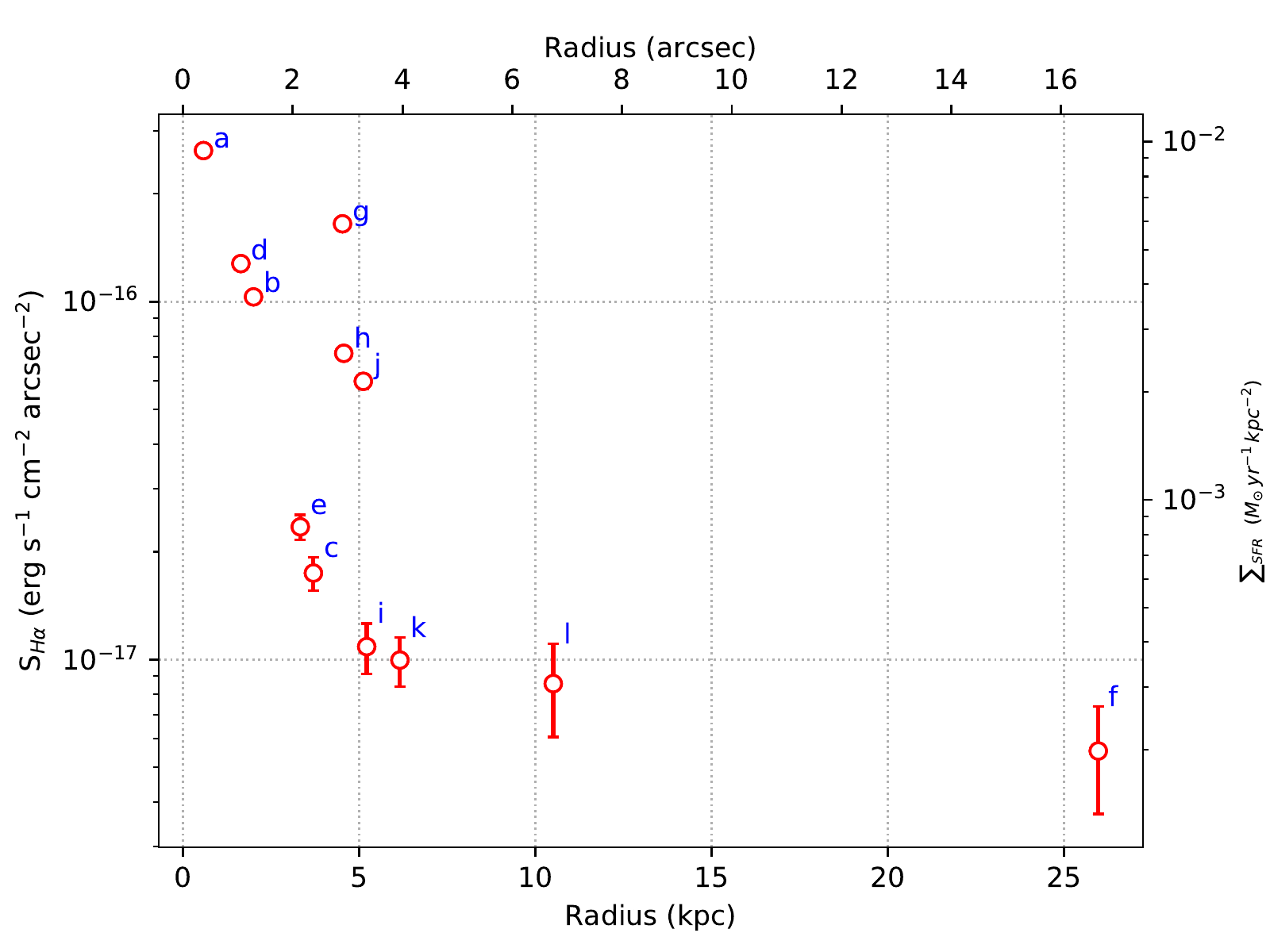}
        \caption{\Ha{} surface brightness measured for different regions of Malin 1. The axis in the right shows the star formation surface density (\sfr{}), corresponding to the observed \Ha{} flux, using the calibration from \citet{boissier13}. The region name of each point is marked as blue letters.}
        \label{surface_brightness_profile}
        \end{figure}

        \figurename{ \ref{surface_brightness_profile}} 
        shows the extracted \Ha{} surface brightness for the 12 detected regions of Malin 1 plotted as a function of the radius.
        There is a steep decrease in the surface brightness for the inner regions of Malin 1, similar to a trend which was observed in the Malin 1 I-Band surface brightness profile by \cite{barth}. This could imply that in the inner regions of Malin 1, the gas profile  follows the stellar profile as in "normal galaxies" \citep{f_combes}.

    The presence of \Ha{} emission in a galaxy is also a direct indicator of star formation activity at recent times (within $\sim$10 Myr; e.g., \citealt{boissier13}), provided there is no other source of ionization like an AGN. However, the effect of a nucleus as a source of ionization is confined to the single central point of our measurements and could not have much effect on our kpc scales (see Appendix \ref{BPT} on the nuclear activity of Malin 1). Therefore, except for the central region, we can convert with confidence the measured \Ha{} flux to the Star Formation Rate (SFR) using standard approximations. We estimate the surface density of star formation rate (\sfr{}) for the 12 regions of Malin 1 following \cite{boissier13} who gives for a \citet{kroupa} IMF:
        \begin{equation}\label{sfr}
                    \text{SFR}\, (M_{\odot}\,yr^{-1}) \,=\, 5.1 \times 10^{-42} \, L_{H\alpha} \,(\text{erg} \,s^{-1})
        \end{equation}
    Our apertures cover several kpc. In the central regions of the galaxy with relatively elevated SFR, we expect to find several \ion{H}{ii} regions per aperture, so that the assumption of quasi-constant star formation history on 10 Myr time-scale for equation \ref{sfr} is valid. In the outer aperture, however, star formation is less elevated and may be stochastic 
    so the derived SFR is less robust.

\section{Discussions}\label{discussion}
    \subsection{The SFR surface density}

\begin{figure}[h]
    \centering
    \includegraphics[width=\hsize]{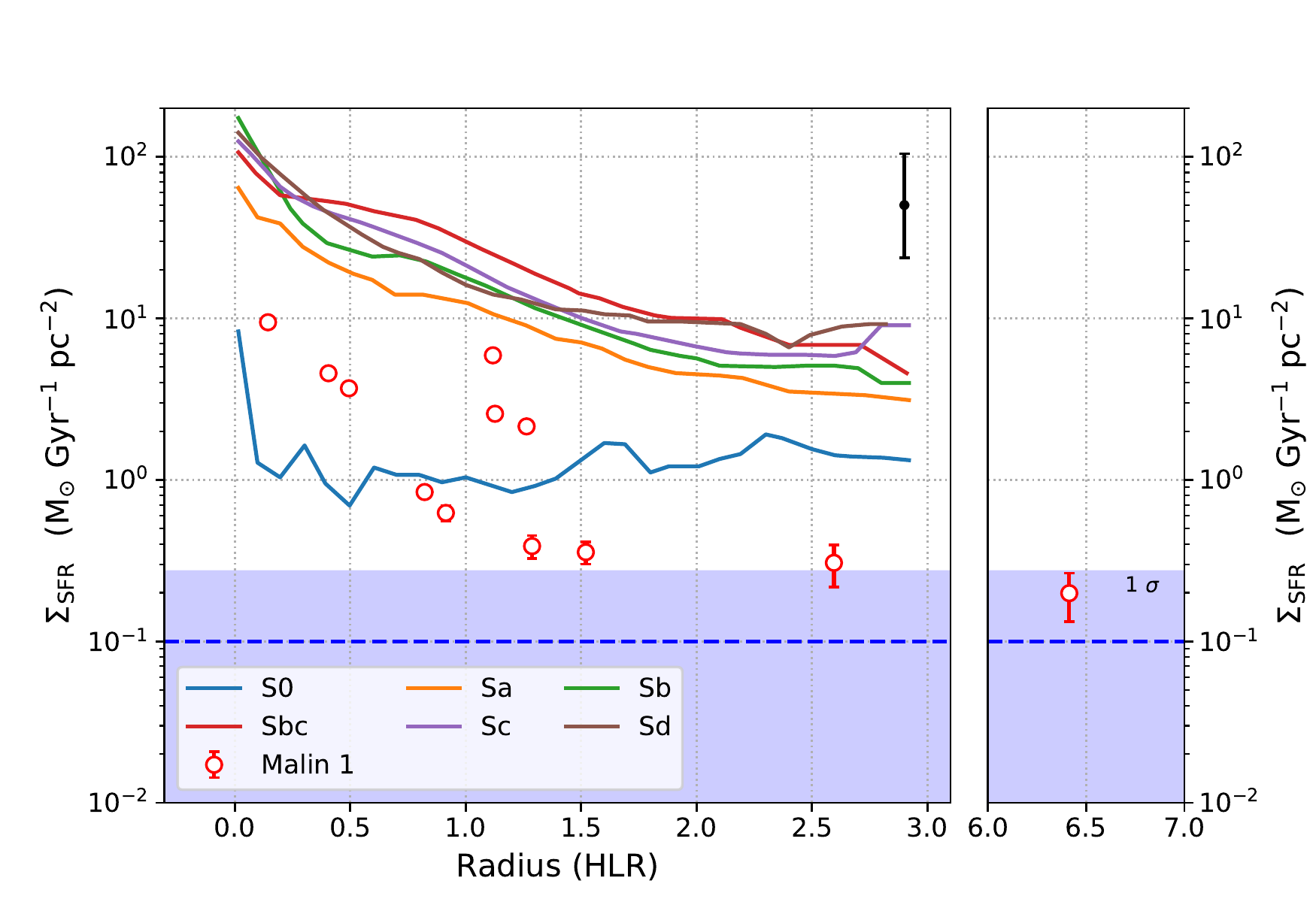}
    \caption{Radial profiles, in units of Half Light Radius (HLR), of the surface density of star formation rate (\sfr{}). The curves correspond to the averages obtained for six morphology of spiral galaxies from \cite{gonzalez16}. The red points shows our extracted \Ha{} data for Malin 1. The blue dashed line indicates the mean level of \sfr{} in the extended disk of spiral galaxies from \cite{bigiel10} with a 1$\sigma$ level of dispersion (blue shaded region). The errorbar in black indicates the typical dispersion among galaxies provided by \citet{gonzalez16} around each solid curve.}
    \label{califa_comparison}
    \end{figure}

    \figurename{ \ref{califa_comparison}} shows a comparison of our Malin 1 estimates of the density of star formation rate at various radius from Sect. \ref{subsecSFRmeasurement} compared to the SFR radial profile for samples of disk galaxies of different morphology from the CALIFA survey \citep{gonzalez16}. These profiles are normalized to the Half Light Radius (HLR). We estimate the HLR of Malin 1 to be equal to  2.6$\arcsec$, calculated within 20$\arcsec$ of the center of the galaxy using the I-band surface brightness profile discussed later and shown in Fig. \ref{I_band_sb}. We adopt this limit so that the comparison is based on the "inner" galaxy at the center of Malin 1 as described by \citet{barth}, not the extended disk, as we believe the CALIFA survey (with data from SDSS) better correspond to this inner galaxy. If we were computing the HLR over the full observed profiles, its value would increase to 18.28\arcsec and the Malin 1 points in Fig. \ref{califa_comparison} would be much more concentrated.

    Our measurements within $\sim$1.5 HLR behave like an intermediate S0/Sa early-type spiral galaxy (\figurename{ \ref{califa_comparison}}), consistent with the observation from \cite{barth}. We have also verified that a comparison with the specific SFR radial profile (using the stellar profile in \citealt{boissier16}) leads to the same conclusion.
    A similar work on two GLSB galaxies Malin 2 and UGC 6614 from Yoachim et al. (in prep., private communication) also shows that GLSB galaxies in general behave like large early type galaxies at their center. 
    Our region at 26 kpc from the center is likely to be part of the extended disk \citep{barth}. 
    Since we detect \Ha{} at only one region really in the extended disk, it is impossible to draw conclusions using this value concerning the overall surface brightness and SFR at that radius. 
    However this surface density of SFR is consistent with the expectations based on the UV blobs luminosity measured in the UV images. It also falls within the 1$\sigma$ dispersion around the average SFR surface density seen in extended disks of spiral galaxies by \citet{bigiel10}, as shown in \figurename{ \ref{califa_comparison}}.
    Finally, a model from \cite{boissier16} for Malin 1 also predicts the SFR surface density around this radius to be $0.08\,\, \text{M}_{\odot}\,\text{Gyr}^{-1}\text{pc}^{-2}$, which is consistent with our measurement.
    However it should be kept in mind that the model predicts the azimuthal average SFR, while our measured value value correspond to a single detected region. 
    Moreover, the detection at 26 kpc is very uncertain due to the sky level. Deeper observations would help to confirm it, and may be detect other faint \ion{H}{ii} region in the extended disk.

    \subsection{Comparison of our rotation curve with other data}
    \label{sec:comparisondata}

    \cite{lelli} provided a rotation curve for Malin 1 using \Hi\ data (\figurename{ \ref{models}}). However the poor spatial resolution of their data makes it hard to study the dynamics in the inner regions of the galaxy ($r$ < 10 kpc) and especially to measure the mass content of the dark matter in LSB galaxies.

     Yet another work on spectral analysis of the inner regions of Malin 1 was performed by \cite{reshetnikov} using stellar absorption lines (shown in \figurename{ \ref{models}}). However, they only provide the radial velocity data corresponding to a single slit position (PA = 55\degr), which we converted to the rotational velocities in the plane of the galaxy taking into account the same geometrical assumptions we have adopted (Sect. \ref{rc_geometry}). The rotational velocities from \cite{reshetnikov} within $\sim10$ kpc are in broad agreement with our data considering their large error bars. This suggests that the gas and the stars rotate together in a coherent way in the central regions of Malin 1. 
     Some stellar absorption lines were also detected in very few regions of our data, which were consistent with our observed rotation curve. However, we prefer not using them considering the poor signal to noise ratio and the small number of detected regions. An accurate comparison of the stellar and gaseous dynamics would require, e.g., IFU data.
     
     In other GLSBs, other behaviors have been sometimes observed, such as counter-rotation in  UGC1922 \citep{saburova2018}.
     Yoachim et al. (in prep., private communication) observes a slowly rising rotation curve for the GLSB galaxies Malin 2 and UGC 6614 using stellar absorption lines, unlike the trend we observe for Malin 1 in this work.

    \subsection{Comparison of our rotation curve with existing models}
    
\begin{figure}[h]
    \centering
    \includegraphics[width=\hsize]{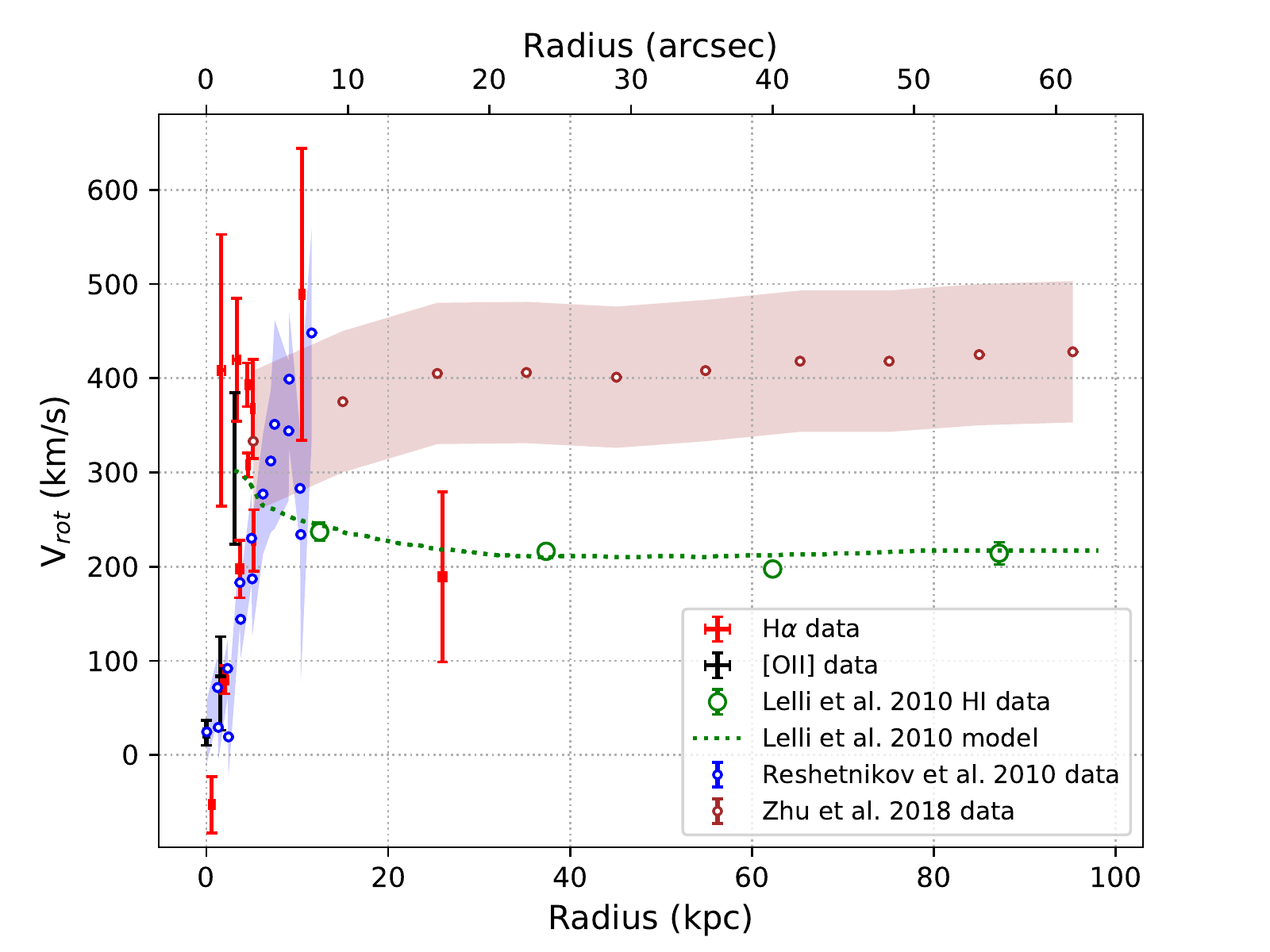}
    \caption{Existing models and data for Malin 1. In addition to the data presented in Fig. \ref{rc}, the green dotted curve is a model from \cite{lelli}, assuming a constant stellar Mass to light ratio (M$_{\star}$/L = 3.4). The brown open circles with the shaded region corresponds to an IllustrisTNG100 simulated data for a Malin 1 like galaxy from \citet{zhu}.The blue open circles are the data from \cite{reshetnikov} using stellar absorption lines with the error bars marked as the blue shaded region.}
    \label{models}
    \end{figure}

     A mass model for Malin 1 from \cite{lelli} using \Hi\ data (shown in \figurename{ \ref{models}}), does not capture the highest rotational velocities we observe for the inner regions and do not show the rise of the rotation curve. This could be due to the low resolution of the \Hi\ data used as the basis of their modeling. 
     Our new observations with high spatial resolution in the center of Malin 1 call for a new mass modeling attempt,
     consistent with our observed rotational curve, taking into account all the stellar, dark matter and gas contributions for the rotational velocity within the galaxy, which will be discussed in the Sect. \ref{mass_model}.

    A recent publication by \cite{zhu} based on the IllustrisTNG simulations also puts forward some interesting results. They were able to find a ``Malin 1 analog'' with similar features to Malin 1 observations and its vast extended low surface brightness disk in the volume of a 100 Mpc box size simulation. They discuss the formation of a Malin 1 analog from the cooling of hot halo gas, triggered by the merger of a pair of intruding galaxies. Their results also include a prediction for the rotation curve of the simulated galaxy with a maximum rotational velocity of 430 km/s (shown in \figurename{ \ref{models}}), close to the maximal value observed for Malin 1 in our analysis. However, the sudden rise of the inner rotation curve for Malin 1 followed by a decline to $\sim$200 km/s seen in our analysis is not observed in their rotation curve. This comparison demonstrates that our observational results offer a new constraint for this type of simulations in the future or any other model of Malin 1 or Malin 1 analogs. Indeed, LSBs and GLSBs can now be studied in the context of cosmological simulations \citep{Kulier2019,Martin2019}.
    
    \section{New Mass Modeling}\label{mass_model}
    We use our \Ha{} and \oii{} rotation curve in combination with \Hi\ measurements from \cite{lelli} and the Hubble Space Telescope (HST) I-band photometry \citep{barth} to construct a new mass model for Malin 1.
    
    We have the total circular velocity components within a disk galaxy given by:
    
    \begin{equation}
        V_{\rm cir} (r)= \sqrt{V_{\rm disk}^{2} + V_{\rm bulge}^{2} + V_{\rm gas}^{2} + V_{\rm halo}^{2}}\\
    \end{equation}
    where $V_{\rm cir}$ is the circular velocity of the galaxy as a function of radius. $V_{\rm disk}$, $V_{\rm bulge}$, $V_{\rm gas}$ and $V_{\rm halo}$ are the stellar disk, stellar bulge, gas and Dark Matter halo velocity components respectively.
    
    We use the \Hi\ gas distribution from \citet{lelli}, corrected for the distance adopted, to derive the gas velocity component.
    However, in order to constrain the stellar bulge and disk velocity components, we need to make a light profile decomposition of Malin 1, discussed in Sect. \ref{sb_decomposition}.

    \subsection{Light profile decomposition}\label{sb_decomposition}
    
    We adopted the I-band light profile provided by \citet{lelli} who combined the HST I-band surface brightness profile of Malin 1 from \cite{barth} for $ r \lesssim$ 10 kpc (high spatial resolution) and the R-band profile from \cite{moore_parker} for $ r \gtrsim$ 10 kpc (large spatial extent). This high spatial resolution in the center is of primordial importance for the rotation curve study, making the use of HST data necessary for the inner part. At larger radii, we checked that this profile is consistent with the recent NGVS \citep{ferrarese} data of Malin 1 \citep{boissier16}. 

    We performed a decomposition of the I-band surface brightness profile following procedures from \citet{barbosa15}, into a S\'ersic bulge, bar and a broken exponential disk component \citep{erwin08} described as:

     \begin{equation}
         I_d(r) = S I_0 e^{-\frac{r}{h_i}}\left[ 1 + e^{\alpha(r-r_{\mathrm{b}})} \right]^{\frac{1}{\alpha}\left(\frac{1}{h_i}-\frac{1}{h_o}\right)}\label{eq_broken_exponential}
     \end{equation}
    
\begin{figure}[h]
    \centering
    \includegraphics[width=\hsize]{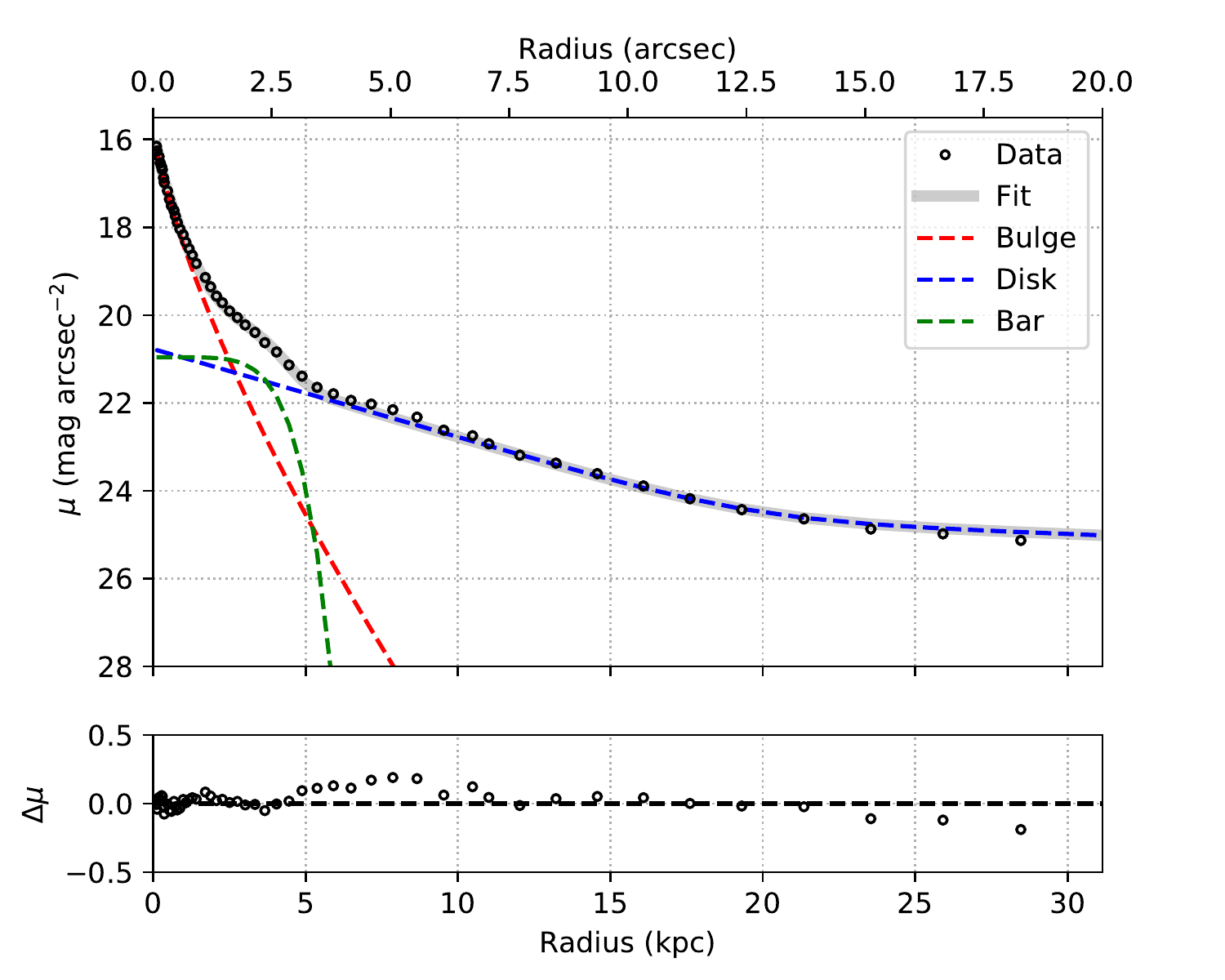}
    \caption{Malin 1 I-band surface brightness decomposition. The bottom panel of the plot represents the difference between the observed surface brightness distribution and the model fit displayed in the top panel.}
    \label{I_band_sb}
    \end{figure}

    The broken exponential function consists of a disk with an inner and outer scale length, $h_{i}$ and $h_{o}$ respectively. The parameters $r_{b}$ is the break radius of the disk and $\alpha$ gives the sharpness of the disk transition. Table \ref{decomp_params} and \figurename{ \ref{I_band_sb}} shows the results of our surface brightness decomposition. It is in good agreement with the decomposition from \citet{barth}, although we obtain a relatively stronger bulge and a weaker bar than in their decomposition. This is a minor difference with negligible effects on our further results. 
    It is well known that bars can cause non-circular motions  \citep{Athanassoula1999,Koda2002,Chemin2015}. However,
    the orientation of the bar in Malin 1 (approximately $45\degr$ with respect to the position angle) could not create major non-circular velocity contributions. Due to the scarcity of measurements, we will not include a bar contribution in our mass modeling.
    Therefore, to make the mass models, we will consider finally two components: the S\'ersic bulge obtained as the fit described previously, and the ``disk'' being the observed profile minus the bulge (in order to account for all the light, but distinguish the spherical geometry of the bulge).
    These profiles are further corrected for the inclination to be used in the mass models.

    \begin{table}[h] 
    \scalebox{1}{
    \begin{tabular}{cccc} 
    \hline \hline \\
    & \multicolumn{3}{c}{\underline{\hspace{3cm}S\'ersic\hspace{3cm}}} \\
     & $\mu_e$ & $r_e$ & $n$ \\
         & (mag arcsec$^{-2}$) & (arcsec) & \\  \hline \\
    Bulge & $18.26\pm0.05$ & $0.59\pm0.05$ & $1.39\pm0.02$  \\
    Bar   & $21.07\pm0.18$ & $1.82\pm0.28$ & $0.17\pm0.36$\\
     \\\end{tabular}}
     
    \scalebox{0.8}{ 
    \begin{tabular}{cccccc}\hline\\
    & \multicolumn{5}{c}{\underline{\hspace{3cm}Broken Exponential\hspace{3cm}}} \\
     & $\mu_0$ & $h_i$ & $h_o$ & $r_b$ & $\alpha$ \\
    & (mag arcsec$^{-2}$) & (arcsec) & (arcsec) & (arcsec) & \\  \hline \\
    Disk & $20.77\pm0.21$ & $3.4\pm0.5$ & $26.9\pm4.9$ & $12.6\pm1.38$ & $0.7\pm0.2$  \\\\
    \hline \hline \\\end{tabular}}

     \caption{Decomposition parameters obtained for Malin 1 from the fitting results. The top two rows show the parameters for the S\'ersic function of the bulge and the bar. The bottom row indicates the parameters for the disk according to the broken exponential function from \citet{erwin08}.}{\label{decomp_params}}
    \end{table}

    \subsection{Mass to Light Ratio}\label{ml_ratio}
    
    During the construction of mass models (Sect. \ref{subsecmassmodels}), the surface brightness are converted into stellar mass profiles, in some cases by fitting the rotation curve, keeping the stellar mass to light ratio as a free parameter. However, it is also possible to "fix" this ratio on the basis of the color index profile.
    \citet{Taylor2011} gives the following empirical relation for the conversion \textit{g}-\textit{i} color to stellar mass to light ratio:
    
    \begin{equation}\label{ML_ color}
        \log(M_{\star}/L_i)_{\odot} = -0.68 + 0.70 (g-i) 
    \end{equation}

    where $M_{\star}/L_{i}$ is the \textit{i}-band stellar mass to light ratio in solar units. \textit{g} and \textit{i} are respectively the \textit{g}-band and \textit{i}-band magnitudes. We use the above relation to obtain $M_{\star}/L_{i}$ to a 1$\sigma$ accuracy of $\approx$ 0.1 dex.

    \begin{figure}[h]
    \centering
    \includegraphics[width=\hsize]{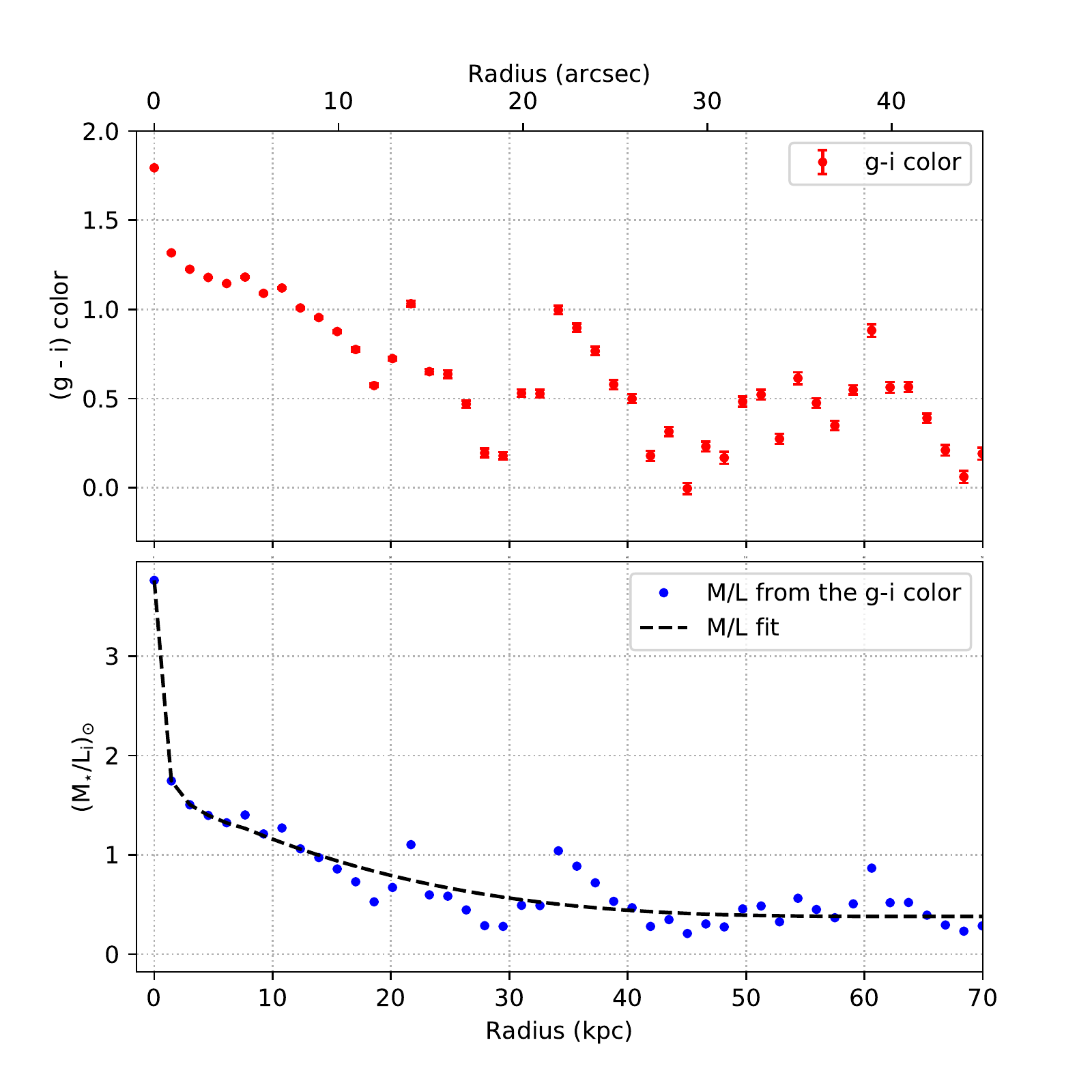}
    \caption{\textit{Top}: \textit{g}-\textit{i} color profile of Malin 1 measured from the NGVS \textit{g}-band and \textit{i}-band images. \textit{Bottom}: Stellar mass to light ratio of Malin 1 in \textit{i}-band measured using the empirical relation from \citet{Taylor2011}. The black dashed line indicates the best fit for the measured $M_{\star}/L_{i}$.}
    \label{color_ML}
    \end{figure}

    We measured a radial profile of the \textit{g}-\textit{i} color from the NGVS images of Malin 1 using the \textit{ellipse} task in IRAF. Our measured values, computed at the NGVS resolution of $\sim$1\arcsec, are in good agreement with the \textit{g}-\textit{i} color of Malin 1 from \citet{boissier16} computed at the GALEX resolution of 5\arcsec. Therefore, we use our measured \textit{g}-\textit{i} color profile to obtain a $M_{\star}/L_{i}$ profile of Malin 1 using the empirical relation given in equation \ref{ML_ color}. Figure \ref{color_ML} shows our extracted color and $M_{\star}/L_{i}$ as a function of radius. 
    We did a polynomial fit on the order of 3 on the extracted $M_{\star}/L_{i}$ profile:
    \begin{equation}\label{ML_radius}
        \frac{M_{\star}}{L_{i}} (r) = 1.69 - 0.0986 r + 0.0025 r^{2} - 0.0000208 r^{3} 
    \end{equation}
    This equation \ref{ML_radius} is only valid for a radius 1\arcsec < $r$ < 40\arcsec. For radius r < 1\arcsec\,, we adopted a peak value of $M_{\star}/L_{i}$ = 3.765 from the color profile. For r > 40\arcsec\,, we adopted a value of $M_{\star}/L_{i}$ = 0.379 in order to make a flat profile for the extended disk.

    In Sect. \ref{subsecmassmodels}, different assumptions are adopted concerning the $M_{\star}/L_{i}$: keeping it as a free parameter, or fix it on the basis of equation  \ref{ML_radius} for the disk, or the  constant $M_{\star}/L_{i}$ value of 3.765 for the bulge.

    \subsection{Beam smearing correction}

\begin{figure}[h]
    \centering
    \includegraphics[width=\hsize]{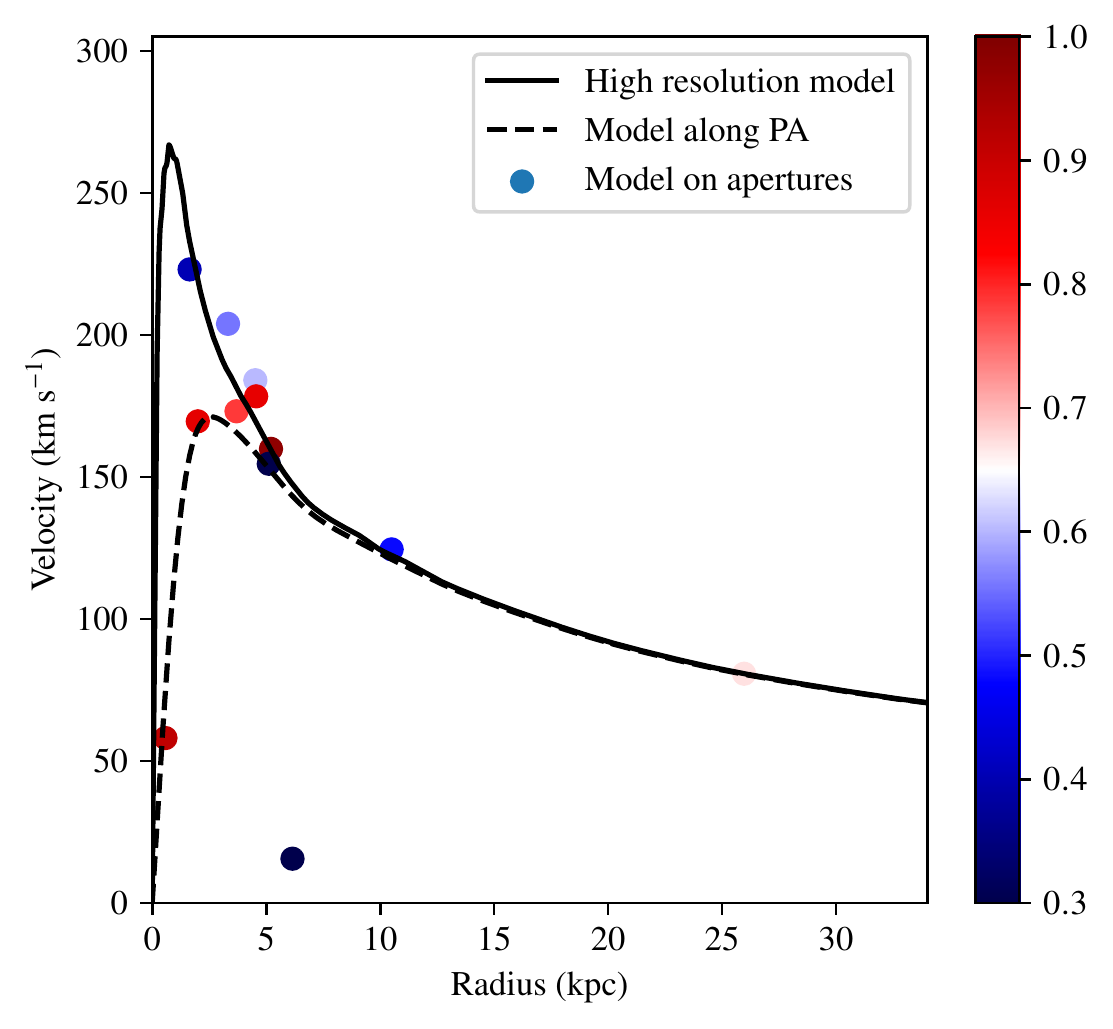}
    \caption{Effect of beam smearing on the rotation curve. The dashed line is the high resolution model, where all the light is supposed to come from the disk, with the M/L that varies as described in Sect. \ref{ml_ratio}; the solid line is the model after accounting for the effect of the beam smearing: the velocity is computed for apertures of 2.5\arcsec\, with a slit aligned with the major axis; the dots correspond to the model on the actual apertures of the \Ha{} dataset and their color indicates their cosine of the azimuth in the galaxy plane.}
    \label{beam_smearing}
    \end{figure}

    The decomposition of the light profile (Fig. \ref{I_band_sb}) is used to compute the circular velocities of the bulge and disk stellar components. We assumed a thin disc and a spherical bulge to compute those velocities.
    They indicate that the rotation is expected to rise more steeply than what is actually observed. One possible reason for this is that the long slit data is severely affected by resolution, due to the seeing, the size of the slit and the apertures used to generate the rotation curve in conjunction with the large distance of Malin 1 and the shape of its inner stellar distribution.
    The impact of this effect can be computed on models, using an observed or modeled light distribution. \citet{Epinat2010} detailed how to perform such computations on velocity fields. In our case we used the expected geometry of Malin 1 (inclination and position of the major axis) together with the light distribution measured in \Ha{} fit by a third order polynomial supposed to be axisymmetric. High resolution velocity fields (oversampling by a factor 8 with respect to the actual pixel size) were drawn from idealized rotation curves derived from stellar mass profiles and the impact of seeing was then modeled using the recipes presented in \citet{Epinat2010}: a convolution of the velocity field weighted by the line flux map normalized by the convolved line flux map. An observational velocity was then derived as the weighted mean of the velocity field on each aperture and the resulting deprojected velocity was computed using the azimuth of the aperture center and the galaxy inclination.
    For each stellar component (bulge and disk), the high resolution mass profiles obtained both with an optimized and varying with radius $M/L$ ratio (see Sect. \ref{ml_ratio}) and without, were used to infer the beam smearing curve.
    The impact of beam smearing is displayed in Fig. \ref{beam_smearing}, which clearly illustrates that beam smearing decreases the amplitude of the velocity and offsets the peak of velocity to larger radii. These modifications depends on both the seeing and the slit width. It also clearly shows that depending on the azimuth, the corrections differ and that it is therefore mandatory to compute the beam smearing for each aperture. In the case of the \Ha{} data, the slit is not aligned with the major axis. Apertures that are centered on the major axis therefore have a different value than the ideal case with the slit aligned with the major axis.
    
    The beam smearing and aperture correction are therefore computed for each individual aperture. These curves were then used in the mass model fitting to describe the stellar components. Such correction was not applied to other components, since they are not expected to strongly dominate in the inner regions where the beam smearing is the most severe.

\subsection{Dark Matter halo}
\label{subsecmassmodels}
To quantify the distribution of the dark matter in Malin 1 we use the observation motivated ISO sphere with a constant central density cored profile \citep{Kent+86}. The core central density profile of the halo is a single power law and the velocity distribution depends on two free parameters, the halo core radius $R_{c}$ and the  velocity dispersion $\sigma$, providing the asymptotical circular velocity as $\sqrt{2} \sigma$:
\begin{equation}
 V_{\rm halo} (r) = \sqrt{2} \sigma \times \sqrt{\left(1 - \frac{R_{c}}{r} \arctan{\frac{r}{R_{c}}}\right)}.
 \label{eq:piso}
\end{equation}
The mass model is adjusted by changing the parameters.

\subsection{Weighting of the rotation curve}
\label{ErrorBars}

The analysis of the mass models depends on the weighting of the rotation curves (RCs), especially when we combine different datasets to construct hybrid RCs as it is the case in this study (\Ha{}, \oii{} and \Hi). The use of a chi-square test ($\chi^2$) for goodness of fit depends on the variance of the independent variables.  Those latters are the rotation velocities and the variance is the uncertainty associated to the rotation velocity estimation. The method for calculating uncertainties may differ, depending on the nature of the data and on the authors. In addition, the density of uncorrelated \Ha\ rotation velocities is in general larger than for \Hi\ RCs and the uncertainties intrinsically larger.
We normalized the uncertainties in attributing the same total weight to the \Ha\ and to the \Hi\ datasets in order to have a similar contribution to the fit from inner and outer regions. Because the weight of uncertainties in a fit is not an absolute but relative quantity, we do not modify the \Ha\ uncertainties but we redistribute the new weights on the \Hi\ data only, using the relation given in \citet{Korsaga+19}.
The weight given to a velocity point is the inverse of its uncertainty. 
The impact of using different weighing methods is discussed in Sect. \ref{sec:resmassmod}.
%

\begin{table*}
\caption{Results of the mass models.
Column (1): Identification of the model.
Column (2): Colors (y: yellow, w: white, OfR: model not plotted because out of the figure range, o: orange, b: blue, lb: light blue, dg: dark green, g: green, l: lime, v: violet, p: pink) and symbols ($\star$ for models using the mass-to-light ratio computed from the color indexes, $\circ$ for model with only one baryonic component, a disk or a bulge) used in Fig. \ref{summary2} to locate the results of the mass models.
Column (3): BFM, MBM and MDM means respectively Best Fit Model, Maximum Bulge Model and Maximum Disk Model. The number after the model indicates the number of free parameters of the fit.
Column (4): Comments on the models. In all the models, except in (c), (d), (e), (i) and (k), we impose the condition \MLd\ $\le$ \MLb; ``c-fix'' means that mass-to-light ratios have been fixed using the color indexes (which is not the case when the mass-to-light ratio is fixed in cases of maximum disk or bulge model); \MLd(r) means that the mass-to-light ratio is a function of the radius as reported in Fig. \ref{color_ML}, while \MLb\ and \MLd\ mean that the mass-to-light ratio is fixed along the radius.
Disk only means that all the stars are distributed in a flat disk component only and Bulge only means that all the stars are in a spherical bulge component only.
Column (5): Best (minimal) reduced chi-square value.
Columns (6 and 7): \MLd\ and \MLb\ are respectively the disk and bulge mass-to-light ratios.
Columns (8 and 9): $R_{c}$ and $\rm \sigma$ are respectively the core radius and velocity dispersion of the dark mater halo.
}
\label{tab:summary1}
\begin{center}
\begin{tabular}{c c c  l r r r  r r}
\hline 
(1)&(2) & (3)& (4)  & (5)&(6)&(7)&(8)&(9)\\ 
 \hline 
 ID&Color& Models& Comments &$\chi_{min}^2$ & \MLd & \MLb & $R_{c}$  & $\rm \sigma$ \\ 
\hline 
&Symb&                      & & & \MLunit & \MLunit & kpc & \kms \\ 
\hline 
(a)&$\ y\ \star$  	& BFM (2) 	& $\rm M/L_{Disk}(r)~\&~\rm M/L_{Bulge}$ c-fix& 12.8  &  1.5--0.5  		    & 3.76   		&  2.4$^{1.1}_{0.3}$  &  142$^{7}_{4}$\\ 
(b)&$\ w\ +$ 		& BFM (4) 	& 	 								& 13.2  &  2.8$^{0.4}_{0.4}$  &  2.8$^{0.6}_{0.1}$  &  2.3$^{0.7}_{0.3}$  &  139$^{3}_{3}$\\
(c)& OfR 	& MDM (\tablefootmark{a} 2/4) 	& $\rm M/L_{Disk} \ge M/L_{Bulge}$        		& \tablefootmark{a} 8.2/9.3  &  20.2$^{0.6}_{0.6}$  &  0.0$^{0.1}_{0.0}$  &  $\infty$  &  0\\
(d)& OfR 	& BFM (4) 	& $\rm M/L_{Disk} \ge M/L_{Bulge}$        		& 7.8  &  15.8$^{1.0}_{0.4}$
  &  0.0$^{0.2}_{0.0}$  &  4.2$^{11.9}_{0.5}$  &  85$^{17}_{2}$\\
(e)&$\ \ \ o\  \square$ & BFM (4) 	& $\rm M/L_{Disk}(r) \ge M/L_{Bulge}$           	& 7.6  &  \tablefootmark{b} 9.7$^{1.0}_{0.3}$  &  0.0$^{0.1}_{0.0}$  &  5.9$^{8.5}_{0.4}$  &  115$^{17}_{2}$\\
(f)&$\ \ \ b\  \star$ 	& BFM (2) 	& $\rm M/L_{Disk}~\&~\rm M/L_{Bulge}$ c-fix	&13.0  &  1.0  &  3.76 &  2.4$^{0.8}_{0.3}$  &  144$^{5}_{3}$\\
(g)&$\ \ \ lb\  \star$ 	& BFM (3) 	& $\rm M/L_{Bulge}$ c-fix 				 	& 12.8  &  3.7$^{0.1}_{0.8}$  &  3.76  &  3.0$^{0.8}_{0.6}$  &  135$^{6}_{1}$\\
(h)&$\ \ \ dg\ \square$ & MBM (3) 	& 									& 16.6  &  2.6$^{2.0}_{0.4}$  &  6.0  &  4.6$^{4.9}_{0.1}$  &  137$^{12}_{4}$\\
(i)&$\ \ g\ \circ$  	& MBM (2) 	& Bulge only 							& 24.8  &  --  	&  9.0  	    	&  15.2$^{0.1}_{8.1}$  &  119$^{2}_{24}$\\
(j)&$\ \ l\ \circ$  		& MDM (2) 	& Disk only 							& 32.2  &  9.0  	&  --  		&  13.0$^{0.3}_{6.0}$  &  121$^{3}_{20}$\\
(k)&$\ \ v\ \circ$  	& BFM (3) 	& Bulge only 							& 12.7  &  --	&  3.1$^{0.6}_{0.2}$  &  2.2$^{1.1}_{0.3}$  &  137$^{4}_{4}$\\
(l)&$\ \ p\ \circ$   	& BFM (3)		& Disk only 							& 13.5  &  2.3$^{0.6}_{0.1}$   	&  --  &  1.9$^{0.8}_{0.1}$  &  142$^{5}_{3}$\\
(m)\tablefootmark{c}& OfR & BFM (2)		& (a)	\& IllustrisTNG100 					& --       &  1.5--0.5  		    & 3.76   		&  3(hs)-4(ls)  &  298(hs)-307(ls)\\ 
(n)\tablefootmark{c}& OfR & BFM (4)		& (b)	\& IllustrisTNG100					& --       &  0(ls)-2(hs)  &  0(ls)-6(hs)  &  2(ls)-5(hs)  &  298(ls)-305(hs)\\
\hline
\end{tabular}
\tablefoot{
\tablefoottext{a}{For models (c), two different degrees of freedom provide two different reduced $\chi^2$ values but identical baryonic and halo parameters.}
\tablefoottext{b}{For model (e), \MLd\ varies with radius as determined from color indexes in Section \ref{ml_ratio} and that the value is the scaling factor with respect to \MLd\ plotted on Fig. \ref{color_ML}, rather than a value in \MLunit.}
\tablefoottext{c}{Models (m) and (n) correspond to the IllustrisTNG100 rotation curve displayed on Fig. \ref{models} instead of the observed one. We consider the same BFM as for models (a) and (b) respectively; (ls) and (hs) in columns (6) to (9) mean ``low slope'' and ``high slope'' respectively, they correspond to the inner slope of the rotation curve until the radius where the IllustrisTNG100 rotation curve begins. No value of $\chi^2$ or uncertainties are given because they depend on the Illustris uncertainties different from the ones in the observations.}}
\end{center}
\end{table*}

\begin{figure}
\begin{center}
\includegraphics[width=8.9cm]{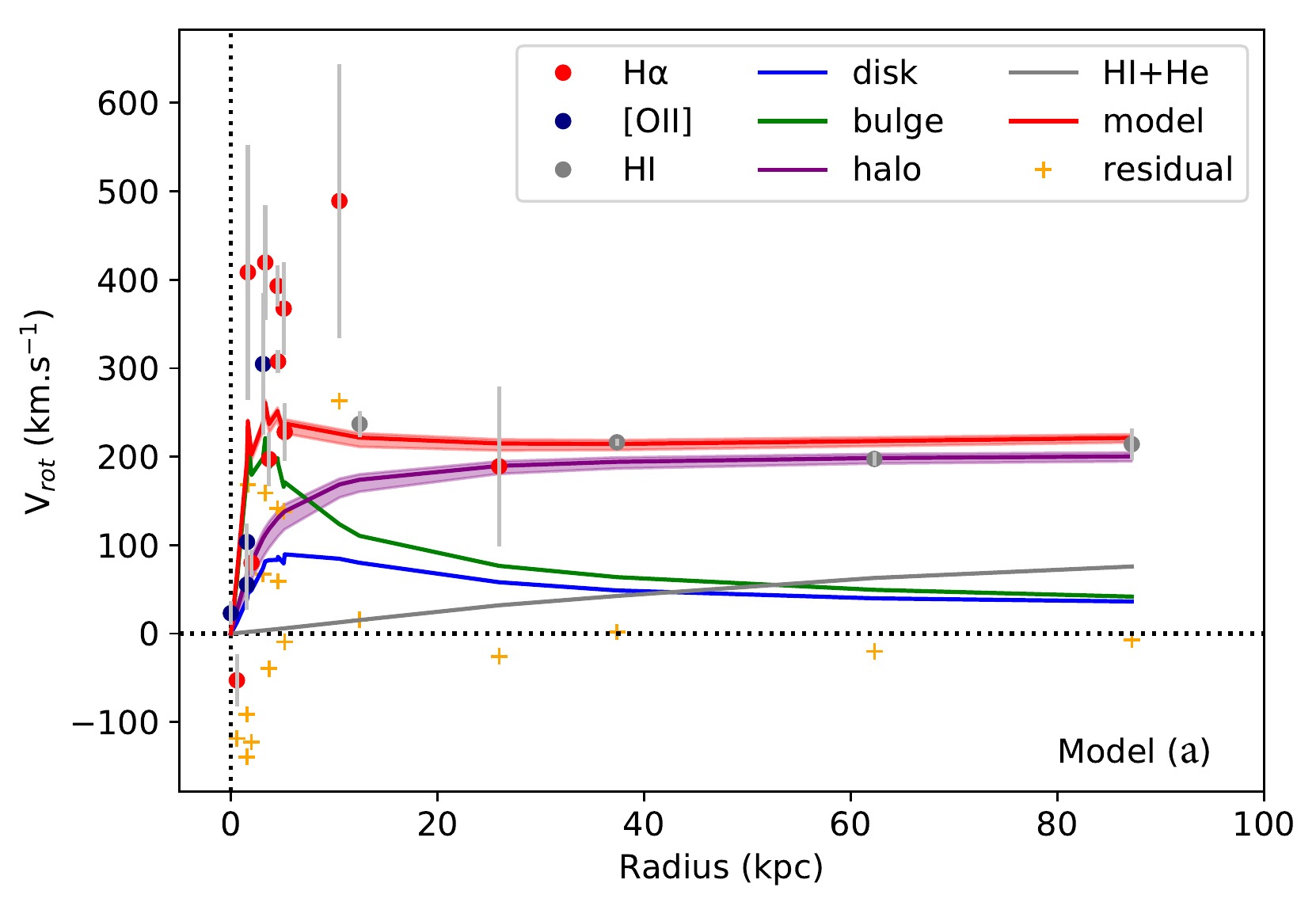}\\
\includegraphics[width=8.9cm]{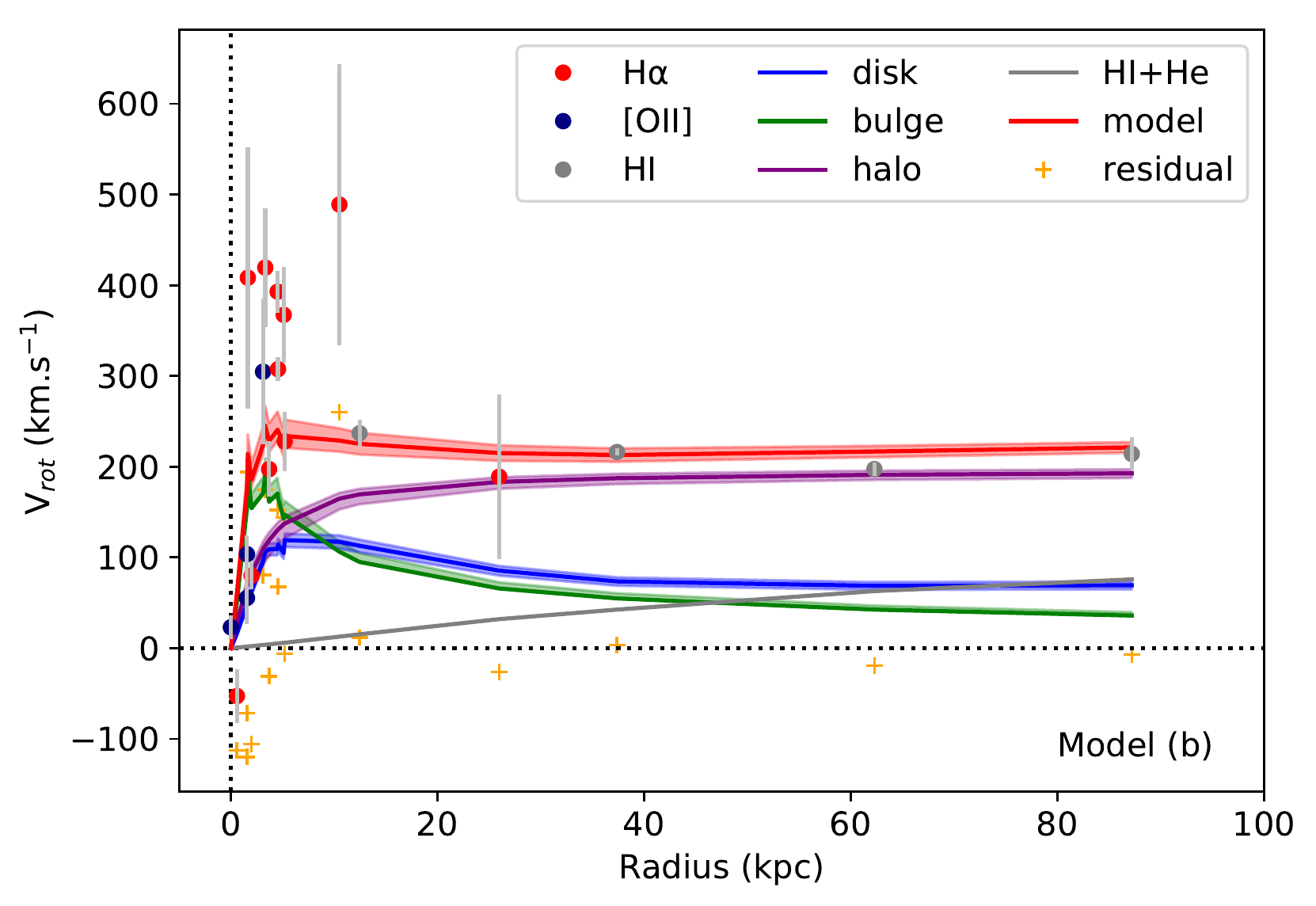}
\end{center}
\caption{
The hybrid rotation curve is plotted using different symbol to represent \Ha{}, \oii{} and \Hi\ data. The resulting model, plotted using a red line, is the quadratic sum of the gas, disk, bulge and dark halo components. The lines correspond to the BFM and the bottom and the top of the filled area around them represent the first and the third percentile around the median (the second percentile) the $\chi^2$ distribution ranging from $\chi^2_{min}$ to 1.10$\chi^2$.
The orange crosses represent the difference between the observed rotation velocities and the model for each point of the RC. Top panel (model a): both disk and bulge mass-to-light ratios are fixed by the color indexes and \MLd\ varies with the radius as displayed in Fig. \ref{color_ML}. The halo parameters are computed using a BFM. Bottom panel (model b): the disk, bulge and halo parameters are fitted using a BFM but \MLd\ is not allowed to be larger than \MLb.
}
\label{summary1}
\end{figure}

\begin{figure*}[ht!]
\begin{center}
\includegraphics[width=\textwidth]{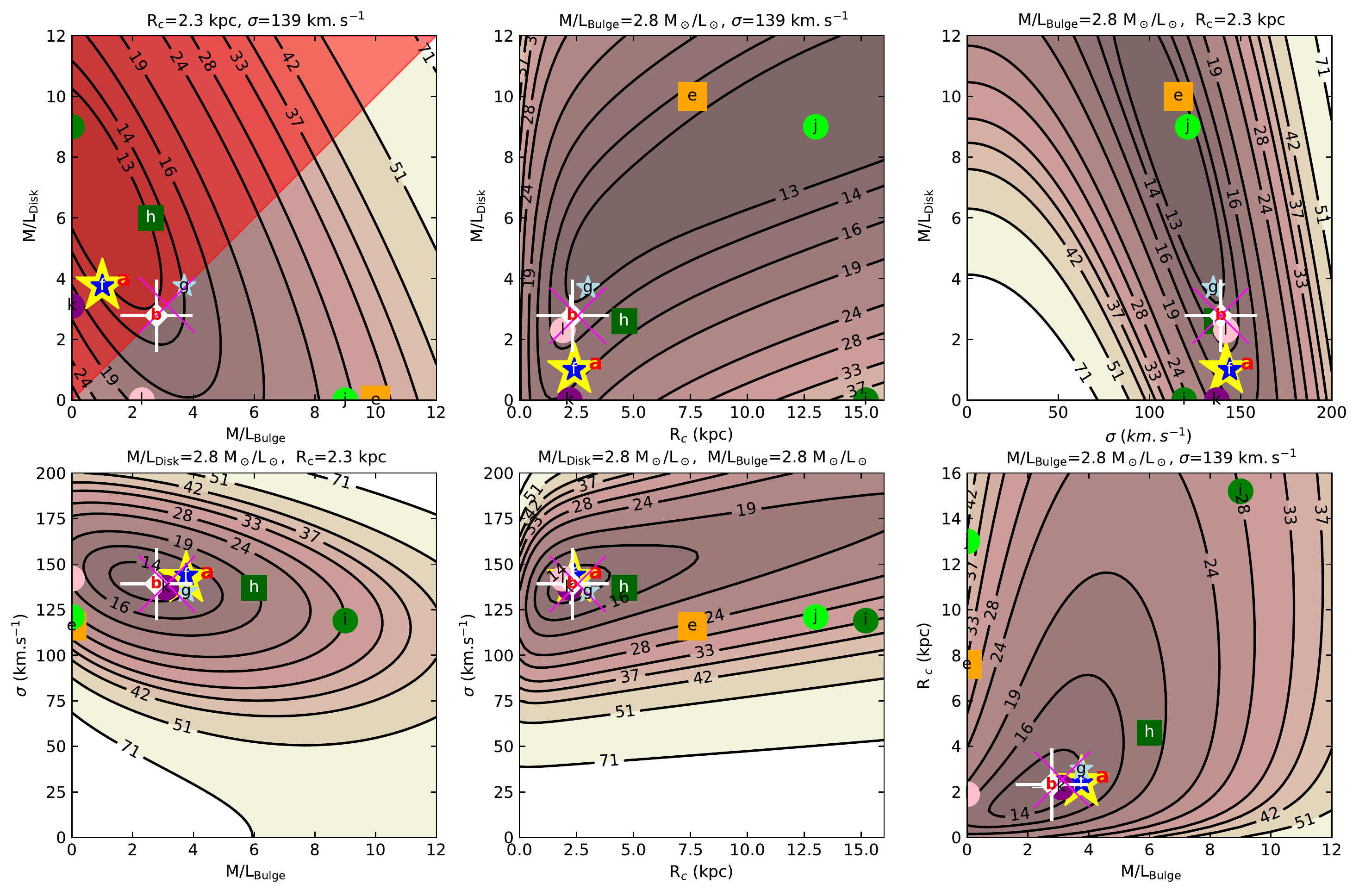}
\end{center}
\caption{
Reduced $\chi^2$ contours for models using mass-to-light ratios determined from the photometry projected on the six planes corresponding to the four-dimensional space \MLd, \MLb, $R_{c}$ \& $\rm \sigma$. Each plane is chosen to match the parameters derived for model (b).
The forbidden area delimited by the condition \MLb\ $\le$ \MLd\ is represented as a red triangle on the top left panel. The contours correspond to the following percentile levels [0, 1, 2, 3, 5, 10, 20, 30, 40, 50, 60, 75, 90] of the distribution, they are labeled by reduced $\chi^2$-values. 
The different symbols, except the large magenta cross,
plotted on the six sub-panels, correspond to ten of the twelve models presented in Table \ref{tab:summary1}, because models (c) and (d) do not fit within the plot limits.
Symbols and their color are described for each model in Table \ref{tab:summary1}. The size of the symbols is random in order to distinguish the different models when they are superimposed.
The large white plus symbols used for model (b), do not match with the isocontour center because of the forbidden volume.
The large magenta cross symbols
correspond to the median of the 5 percentile of the $\chi^2$ distribution.
}
\label{summary2}
\end{figure*}

\subsection{Results of the mass modeling}\label{sec:resmassmod}

We present in Fig. \ref{summary1} the mass model for two cases. On the top panel (model a) the disk and the bulge stellar mass-to-light ratio (\MLb) have been computed using the color indexes as discussed in Sect. \ref{ml_ratio}, and the disk one (\MLd) varies with the radius as shown in Fig. \ref{color_ML}. The core radius $R_{c}$ and the velocity dispersion $\sigma$ of the dark matter halo have been adjusted using a best fit model (BFM). On the bottom panel (model b), the four parameters of the model are let free to vary, they are minimized by a BFM, and \MLb\ is also forced to be larger than \MLd. The parameters of those two models are given in Table \ref{tab:summary1}
Those two models lead to similar halo parameters despite the fact that the disk component is about twice larger in model (b) than in model (a) because the larger disk is almost compensated by a weaker bulge in model (b). The dark matter halo dominates the baryonic components from $\sim$10 kpc to the end of the RC at $\sim$100 kpc and the bulge is requested to fit the RC within the first five kpc, the dark matter halo being not cuspy enough. The main issue of both models is, however, that none is able to fit correctly the center of the RC and rotation velocities larger than $\sim$250 \kms.\\

We tried the ability of other models to fit these inner points of the RC and report the results for some of those in Table \ref{tab:summary1}.
Model (c) is a MDM where we force the halo to vanish and the disk to be maximal in order to better fit the inner velocity points of the RC. The $\chi^2\sim$8.2 is smaller than for models (a) and (b) meaning that the fit is better on average. This set of parameters is fully compatible with no bulge and no halo but the price to pay for that is an unrealistically large \MLd\ $\sim$20.2 \MLunit, compared to \MLd $\sim 1 \pm 0.5$ \MLunit\ provided by color indexes. Because this model has no halo, it only implies two free parameters (the baryonic components). However, if we divide the non-reduced $\chi^2$ by the number of data points (19) minus 4 free-parameters instead of 2, we get a reduced $\chi^2 \sim 9.3$ as reported in Table \ref{tab:summary1}, columns (3) and (5).
Model (d), which is a BFM with four free parameters allowing \MLd\ to be larger than \MLb, provides similar results as model (c) with the same absence of bulge, a strong disk and a weak halo, despite it shows that the disk alone cannot adjust the RC.
Models (c) and (d) hardly fit some inner velocity points around 300 \kms, they cannot fit either the highest ones, and, in addition, rotation velocities between 10 and 30 kpc are largely overestimated.
We conclude that the ``natural'' solution when \MLd\ can (non-physically) overpass \MLb, is a model without DM halo and without bulge (or very marginal halos and/or bulges).
It is interesting to compare the mass models when the mass-to-light ratio does not changes or does change with the radius.
Indeed, in this latter case, it varies from $\sim$2 \MLunit\ in the center of the galaxy, to $\sim$0.38 \MLunit\ at a radius $\sim$70 kpc, with a median value of $\sim 1$ within the first 25 kpc, where most of the velocity measurements are and where the disk contribution is the highest (see Fig. \ref{color_ML} and Section \ref{ml_ratio}).
Best fit models (d) and (f) used a \MLd\ that does not vary with the radius while it does in models (a) and (e).
By definition, the best $\chi^2$ parameter is obtained when all the parameters are fully free to optimize the fit, without any constraints. This is the case of models (d) and (e) for which the physical constraint \MLb\ $>$ \MLd\ is not imposed to the fit, thus we get the smallest $\chi^2 \sim 7.7$, almost 1.7 smaller than for models (a) and (b), which provide $\chi^2\sim$13.
Comparing models (d) and (e) shows that the average \MLd\ of model (d) is $\sim 1.5 \times$ larger than the one of model (e). The consequence is a stronger halo in model (e) to compensate for its weaker \MLd\ at large radius.
Model (f) is a variation of model (a) for which the \MLd\ does not change with the radius. The results for models (a) and (f) are very similar. The impact on the halo seen in model (e) with respect to model (d) is not observed anymore because  the bulge contribution is similar to that of the disk.
In order to check if the inner part could be better fitted, we release the constraint on the \MLd\ fixed by color indexes in model (g) but we keep the one on the bulge. As expected, \MLd\ tends to increase almost to the value of \MLb, the growing of the disk has for consequence to marginally weaken the dark matter halo but not to better fit the inner velocity rotations and both models (f) and (g) provide almost the same $\chi^2$ ($\sim$13.0 and $\sim$12.8 respectively).
To step further, because the bulge shape is sharper in the inner region than that of the disk, in order to try to fit the very inner points, we maximize \MLb\ to 6 \MLunit\ in model (h). This large, but still reasonable \MLb\ value (less than twice as high as the one determined using color indexes) allows the model to pass through the large bulk of dispersed rotation velocities within the 5 inner kpc, but therefore does not allow to reach the highest rotation velocities around 400 \kms. In addition to the fact that this model does not reach the velocities above 300 \kms, it poorly fits the velocities within 0 and 5-10 kpc.
We note that \MLd\ and DM parameters are highly degenerated in model (h) since a high \MLd\ and a weak DM halo provide almost the same $\chi^2$-value than no disk and a stronger DM halo.
However, the principal issue of this model is the slope of the bulge that is far too sharp to fit the rotation velocities around 200 \kms\ but has astonishingly the right shape to fit the velocity around 400 \kms\ even though its cannot reach them. We clearly see two regimes of velocities at the same radius within the first 5 kpc. The poor fit within the first 5 kpc of model (h) provides a high $\chi^2\sim 20$ despite the fact that it fits fairly well the velocities at larger radius, between 10 and 100 kpc. In the last four models, from (i) to (l), we study the impact of the stellar distribution geometry on the fits, since this modifies the shape of the model rotation curve.
We observe a bright and peaked surface brightness distribution in the center which could be either a spherical bulge or eventually a bright nucleus. In models (i) and (k) we therefore distribute all the stars in a spherical bulge component while in models (j) and (l) we set it in a flat disk component. For models (i) and (j) we maximize the stellar components using for both the same fixed mass-to-light ratio, while in models (k) and (l) we test a BFM to let the halo take place. 
None of those 4 models allows to describe better than the others the mass distribution of Malin 1 in the sense that BFM models provide almost the same $\chi^2$-values but do not help to fit the highest rotation velocities, and if maximum disk or bulge models permit to better reach them, they increase the velocity dispersion of residuals velocities, which is also indicated by their high $\chi^2$-values.\\

{Models (m) and (n) correspond to the IllustrisTNG100 RC displayed on Fig. \ref{models} and discussed is section \ref{sec:comparisondata}.  We consider the same BFM as for models (a) and (b) respectively, except that the observed RC is replaced by the IllustrisTNG100 one, in which we add arbitrary rotation velocities in the first five kpc of the galaxy. Indeed, the IllustrisTNG100 RC starts only at a radius of $\sim$5 kpc while the observed light surface brightness provides constraints within this radius, which are used in the present analysis.  We define two different inner slopes for the rotation curve within the first point of the IllustrisTNG100 RC (at a radius of $\sim$5 kpc).  In table 5, (ls) refers to a ‘‘low slope'', more precisely an almost solid body shape from 0 to 5 kpc with a velocity gradient of $\sim$65 \kms\ kpc$^{-1}$,  while (hs) indicates a ‘‘high slope'',  more precisely a solid body shape RC with two slopes, with a velocity gradient of $\sim$130 \kms\ kpc$^{-1}$ from 0 to 2.5 kpc and a lower slope from 2.5 to 5 kpc to smoothly join the IllustrisTNG100 RC. As a consequence of the fact that the IllustrisTNG100 RC amplitude at large radius is twice as large as the observed RC, for both models (m) and (n), the asymptotical velocity dispersion $\sigma$ is also about twice as large as for previous models, where the actual rotation curve is used.  $\sigma$ is almost the same ($\sim$ 300\kms) for models (m) and (n), it does not depend on the inner slope of the RC and on the fact that disk and bulge components are used or not.  At the opposite, in model (b), where all parameters are free, the inner slope has a direct impact on the bulge and disk mass-to-light ratios: a low inner velocity gradient provides a mass model in which no baryonic component is requested, while a high inner velocity gradient allows significant baryonic components.
In model (m), for which bulge and disk components are fixed by the colors, the core radius tends to match the inner slope of the RC: it is slightly larger/smaller in the case of a low/high inner slope. For model (n), in the case of a high inner RC slope, a high bulge mass-to-light ratio enables to fit the inner velocities, thus the core radius is more than twice as large than in the case of a low slope where no bulge can fit.  The halo mass estimated at the last \Hi\ radius (87.2 kpc), using relation \ref{eq:piso}, gives almost the same value of $3.46\pm 0.05\times 10^{12}$ M$_\odot$ for models (m) and (n) and for the two different inner slopes (ls and hs).   This is due to the fact that the total baryonic mass and the baryonic matter distribution do not strongly affect the halo shape.  In conclusion, the IllustrisTNG100 RC does not reproduce the different datasets, neither at large nor at small radius and no constraint is given on the baryonic components when the RC does not provide any constraint in the inner regions.}\\

In order to compare the different models described in this section and tabulated in Table \ref{tab:summary1}, we plot on Fig. \ref{summary2}, the parameters of the different mass models. The isocontours of all the panels involving $\sigma$ show that this is the best constrained parameter: all the dark matter halo reach a similar asymptotical velocity around $130 \sqrt{2}$ \kms. In addition, most of the dark matter halos are relatively concentrated with a core radius $R_{c}\sim$2.5 kpc, this confirms that a massive dark matter halo is mandatory to adjust the observations. As shown by the different model and the isocontours of i.e. panel 5 ($R_{c}$ vs \MLd), the natural trend of the disk is to be massive to extremely massive (for \MLd\ $\sim 6-26$\MLunit) but this is an artifact linked to the lack of velocity resolution in the inner part. \MLb\ tends to be compatible with the one determined by the color index method if \MLb\ value is not allowed to overpass \MLb\ one.\\

As introduced in the Sect. \ref{ErrorBars}, the weighting of the rotation velocity coming from \Hi\ and optical datasets can differ significantly. In the present study, among the 19 independent velocities measurements, only four come from \Hi\ data. In addition, the mean uncertainty are ten time larger on the optical dataset than on the \Hi\ one ($\sim$53 and $\sim$5 \kms\ respectively). Because of the importance of the weighting definition, we test the impact of the uncertainties on the model by testing additional methods to weight the data.
We presented a first method in Sect. \ref{ErrorBars}, a second method simply consists of using the original error bars, i.e. the ones computed independently for the optical and the radio datasets. In that case, the output parameters are only marginally modified, this is mainly due to the fact that each \Hi\ error bar is naturally typically ten times smaller than an optical uncertainty, thus providing on average a weight ten times larger than the optical one, compensating the fact that only $\sim$1/5 velocity measurements come from \Hi\ data.
In a third method, we give the same weight to all the velocities, independently of their wavelength or their radial and azimutal locations in the galaxy. This weight is taken as the median of the uncertainties of all the velocities. 
In that case, the weight of the outer \Hi\ data is on average five times smaller than the one of the optical data and the mass models tend to be poorly constrained in the outer region, i.e. for radius larger than 30-40 kpc, which is not really acceptable with respect to the large size of Malin 1. 
Thus rather than going forward with this third method in which the outer \Hi\ velocity points have a weak weight, we prefer to discuss the case study for which no \Hi\ velocity point is used, but where the \Hi\ surface density contribution is kept. In this fourth configuration for the uncertainties, we only fit the $\sim$30 inner percents of the RC.
When the \Hi\ velocities are not used, the disk mass-to-light ratio decreases on average by $\sim$38\% when it is a free parameter (models b, c, d, e, g, h, l, see Table \ref{tab:summary1}), and the bulge mass-to-light ratio decreases from an average value $\sim$1.2 to 0 \MLunit\ when it is a free parameter (models b, c, d, e, k).
Except for case (c), the halo parameters are allowed to freely vary.
Using the RCs without the \Hi\ components, the halo disappears for cases (i) and (j).
In order to compare the behavior of the core radius $R_{c}$ that becomes infinite when no halo component is involved, cases (c), (i) and (j) are furthermore discarded.
The mean $R_{c}$ increases by $\sim$68\% from 3.2 to 5.4 kpc, and the mean velocity dispersion $\sigma$ grows from $\sim$129 to $\sim$260\kms\, i.e. increases by $\sim$124\%\footnote{We do not allow the halo asymptotical velocity $\sqrt{2} \sigma$ to reach a value larger than the maximum observed rotation velocity $V_{Max}=498$ \kms.
We therefore impose an upper limit $\sigma_{Max}=346$ \kms\ to the velocity dispersion.
Except in cases (c), (i) and (j), where $\sigma=0$ \kms, $\sigma=\sigma_{Max}$, which leads to the average value $\sigma=260$ \kms.
}.
Those trends mean that baryonic components are largely when no \Hi\ velocities are taken into account while the dark halos are less concentrated but reach larger velocity dispersions.  In terms of mass, using relation \ref{eq:piso}, when the whole RC is used, the mean halo mass measured at the last \Hi\ radius (87.2 kpc) is $5.7\pm 2.3\times 10^{11}$ M$_\odot$ while it inconsiderately jumps to $3.3\pm 1.9\times 10^{12}$ M$_\odot$ without \Hi\ velocities, i.e. an increase by a factor $\sim$5.7. \\\indent
On the other hand, the average halo mass computed without the \Hi\ velocity for models (a) and (b) is $\sim 4.3\times 10^{12}$ M$_\odot$, which is only $\sim$1.25 times larger that the halo mass estimated using the IllustrisTNG100 RC for the same models ($\sim 3.5\times 10^{12}$ M$_\odot$). This means that if we only had the \Ha\ RC, we would have thought that the IllustrisTNG100 RC gave a compatible model at large radius.\\
\indent To conclude this discussion, a final test has been done. Indeed, looking to the observed RC, we note that the \Ha\ rotation velocities strongly decrease between the two largest radii from $V_{Max}\sim$489 \kms\ (at a galacto-centric radius of $\sim$10.5 kpc) to $\sim$189 \kms\ (at $\sim$26.0 kpc), i.e. to a rotation velocity lower than the average \Hi\ velocities of $\sim$216 \kms.  In order to check that this outermost \Ha\ velocity does not bias the results when we only use the optical data, we rerun the previously called ``fourth configuration'' without considering it. This still increases the disk mass-to-light ratio by $\sim$31\% (instead of $\sim$38\%), and decreases the ratio between the mean halo mass with and without \Hi\ from 5.7 to 5.2.  Without this outermost \Ha\ velocity, the constraints are still released,
however, the disk mass-to-light ratio and the halo parameters do not change much, this outermost \Ha\ data point being weighted by the other 18 optical measurements.

In summary, the new high resolution and extended rotation curve of Malin 1 seems to not require a strong dark matter halo component in the inner part, where the the observed stellar mass distribution can explain the dynamics, but a massive dark matter halo remains necessary to fit the outer regions, whatever the mass model performed.
The fit in the inner parts is poor and this may be related to the spatial resolution of the observations that is at the very limit not to be impacted by beam smearing but also to the assumptions made on the geometry of the inner gaseous disk. Ideally, high spectral ($R>2000$) and spatial (PSF FWHM $<1$'') resolution integral field spectroscopy data could solve these discrepancies by providing at the same time the actual ionized gas line flux spatial distribution and the geometry of the gas in these inner regions, thanks to their 2D kinematics, with much smaller uncertainties.
A discussion on the shape of the dark matter halo (cusp vs core) is therefore out of the scope of this paper.

\section{Conclusions}\label{conclusion}

We present a spectroscopic study of the GLSB galaxy Malin 1 using long-slit data from IMACS spectrograph. In this work we focussed on the \Ha{} and \oii{} emission lines detected in 16 different regions of Malin (12 \Ha{} and 4 \oii{} detections). Key results of this work are :
\begin{itemize}
\item We extracted a new rotation curve for Malin 1 using \Ha{} and \oii{} emission lines, up to a radial extend of $\sim$26 kpc.
\item For the first time we observe a steep rise in the inner rotation curve of Malin 1 (within r < 10 kpc), which is not typical for a GLSB or LSB galaxy in general, with points reaching up to at least 350 km/s (with a large dispersion) at a few kpc before going back to 200 km/s, the value found in \Hi\ at low resolution.
\item We made an estimate of the \Ha{} surface brightness and star formation rate surface density of Malin 1 as a function of radius, using the observed \Ha{} emission line flux. The \sfr{} within the inner regions of Malin 1 is consistent with an S0/Sa early type spiral. The region detected at $\sim$26 kpc from the center of Malin 1 has a \sfr{} close to the level found in the extended disk of spiral galaxies.
\item An analysis of the observed Balmer line ratio indicates a very low amount of dust attenuation within Malin 1 (consistently with previous works in the infrared).
\item Line ratios, however, point to a relatively high metallicity for the inner regions.
The line ratios in the center are consistent with the previous classification of Malin 1 as a LINER/Seyfert (see Appendix \ref{BPT}).
\item The new high resolution and extended rotation curve of Malin 1 seems to not require a strong dark matter halo component in the inner part. In these regions, the observed stellar mass distribution can explain the observed dynamics. However, a massive dark matter halo is required in the outer regions.
\item The fit of the rotation curve in the inner parts is poor. This may be due to the coarse spatial resolution of the observations, but also to the assumed geometry of the inner gaseous disk (e.g., non-circular velocity contributions due to the bar).
\end{itemize}

This work allows us to provide new constraints on Malin 1. 
It will be important in the future, however, to obtain better quality and complementary data for Malin 1, as well as for other giant LSBs, e.g., with optical IFU such as MUSE or with ALMA, to bring more constraints on the origin of these galaxies. In the recent years, it became possible to obtain deeper observations with new telescopes such as Dragonfly \citep{dragonfly} or new instrumentation (e.g., MegaCam at CFHT, Hyper Suprime-Cam at Subaru, Dark Energy Survey Camera at the 4m Blanco telescope). This allowed astronomers to re-visit the low surface brightness universe, including GLSB galaxies \citep[e.g.,][]{galaz,boissier16, hagen}, extended UV galaxies discovered with GALEX \citep{gil_de_paz,thilker05} and even to define and study the new class of ultra-diffuse galaxies (UDGs) \citep[e.g.,][]{van_dokkum,koda15}. Galaxies like Malin 1 will be detectable, if they exist, up to redshift 1 with upcoming projects like SKA \citep{ska}. The recent studies of Malin 1 and other LSBs or UDGs shows that the low surface brightness universe has a \emph{bright} future.\\

\begin{acknowledgements}
We acknowledge the support by the Programme National Cosmology et Galaxies (PNCG) of CNRS/INSU with INP and IN2P3, co-funded by CEA and CNES. We thank M. Fossati for providing some plotting codes for the line diagnostic diagram and S. Arnouts for his help on the environment of Malin 1. This work also utilised some data from the SDSS DR12 Science Archive Server (SAS) for comparison purposes.
Funding for SDSS-III has been provided by the Alfred P. Sloan Foundation, the Participating Institutions, the National Science Foundation, and the U.S. Department of Energy Office of Science. The SDSS-III web site is http://www.sdss3.org/.
\end{acknowledgements}

\bibliographystyle{aa}
\bibliography{bibliography.bib}

%

\begin{appendix}

\section{Estimation of errors}\label{appendix_error_estimation}
There are mainly two sources of errors in the data provided in this work (see Table. \ref{data}) - the emission line fitting error and the slit positioning error. We calculated these errors separately as detailed below. They are combined and propagated to obtain the final errors in each of our quantities of interest.

\subsection{Emission line fitting error}

    The emission line fitting of the spectra (both \Ha{} and \oii{} data) was carried out using an initial formal fitting followed by a Monte Carlo Markov Chain (MCMC) method to obtain the final fitting results. The formal fitting of the spectrum was done using the \textit{scipy.optimize.leastsq} python package (see Fig. \ref{fittting_results} for fits of continuum + Gaussian emission lines, from which are derived line positions, intensities, line width). From this fit, a model spectra is obtained from the sum of the different component (continuum and lines): $F_{\mathrm{model}}(\lambda)$.

    The noise standard deviation $\sigma_{\mathrm{noise}}$ was estimated by subtracting $F_{\mathrm{model}}$ from the observed spectrum ($F_{\mathrm{obs}}$) in order to obtain the residual spectrum. In order to remove any wavelength dependence, this spectrum was fitted with a polynomial of order 3 that was subtracted to it. We then measured the statistics of this flattened residual spectrum in order to obtain $\sigma_{\mathrm{noise}}$. 

\begin{figure}[h]
        \centering
        \includegraphics[width=\hsize]{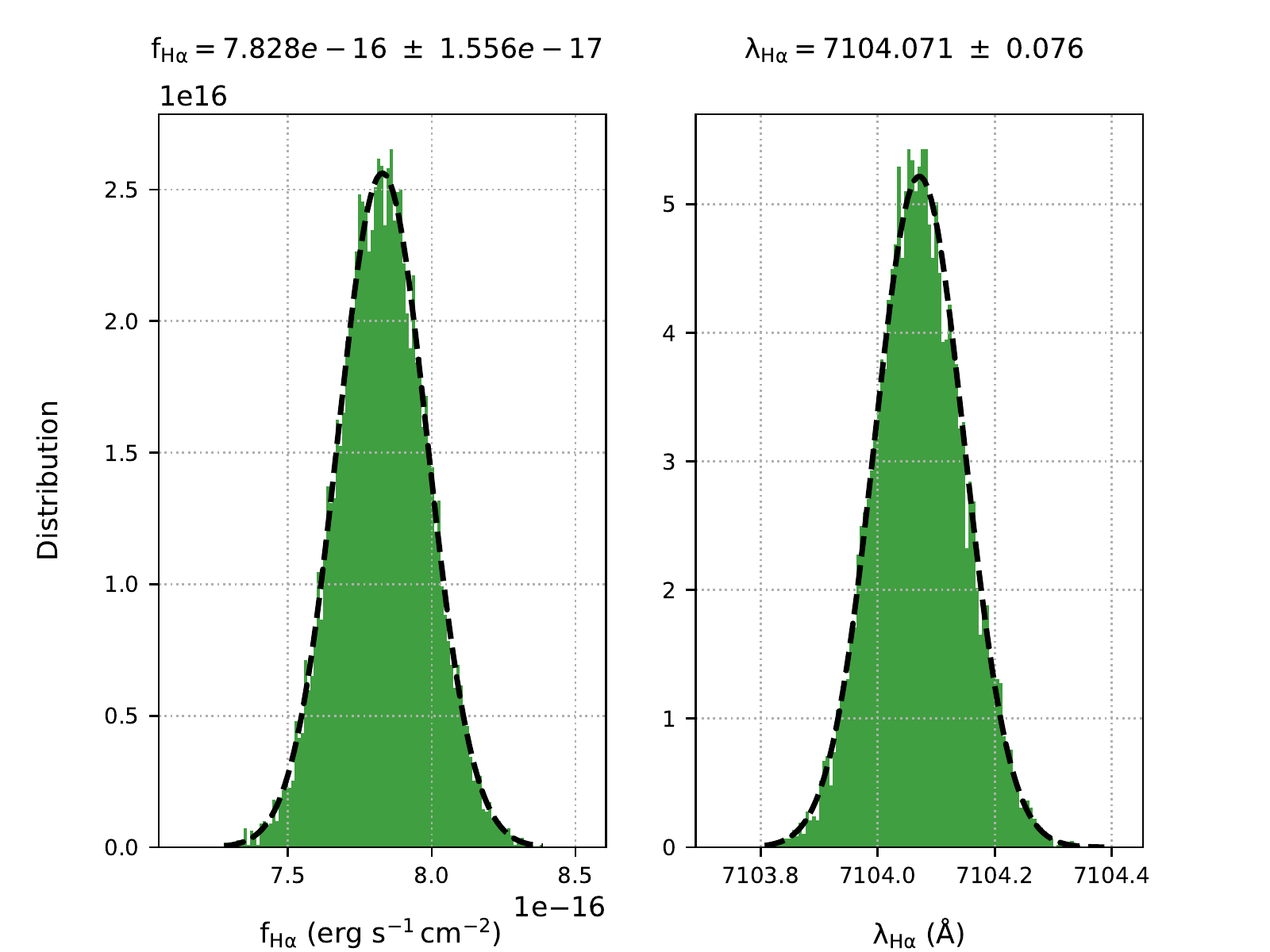}
        \caption{A example of the MCMC fitting results for the \Ha{} line in the region \textit{a} (see Table. \ref{data}). The left and right panels respectively shows the flux distribution ($f_{H\alpha}$) and peak wavelength positions ($\lambda_{H\alpha}$) from the MCMC iterations. The mean value of this quantities along with their $1\sigma$ errorbars are labeled on top of each panel.}
        \label{mcmc_histogram}
        \end{figure}

    We then performed an MCMC fitting procedure using an iterative chain of $ N = 10000$ iterations. In the beginning of each iteration, a synthetic spectrum $F_{\mathrm{syn}}$ was created using the $F_{\mathrm{model}}$ with the addition of a random noise with standard deviation $\sigma_{\mathrm{noise}}$. 
    At each iteration,  we performed again a formal fitting of $F_{\mathrm{syn}}$, providing in each case a mock determination of the amplitudes, peak wavelengths, widths and continuum level of all the emission line components in the fit.
    Then histograms of these determinations for each parameters (we focussed on the emission line peak wavelength and flux) were created, inspected, and fitted with a normal distribution. The mean value and standard deviation of each histogram distribution respectively gives our best fit value and fitting error of a parameter (see Fig. \ref{mcmc_histogram} for an example).
    
\subsection{Slit positioning error}\label{section_slit_pos_error}

\begin{figure}[h]
    \centering
    \includegraphics[width=\hsize]{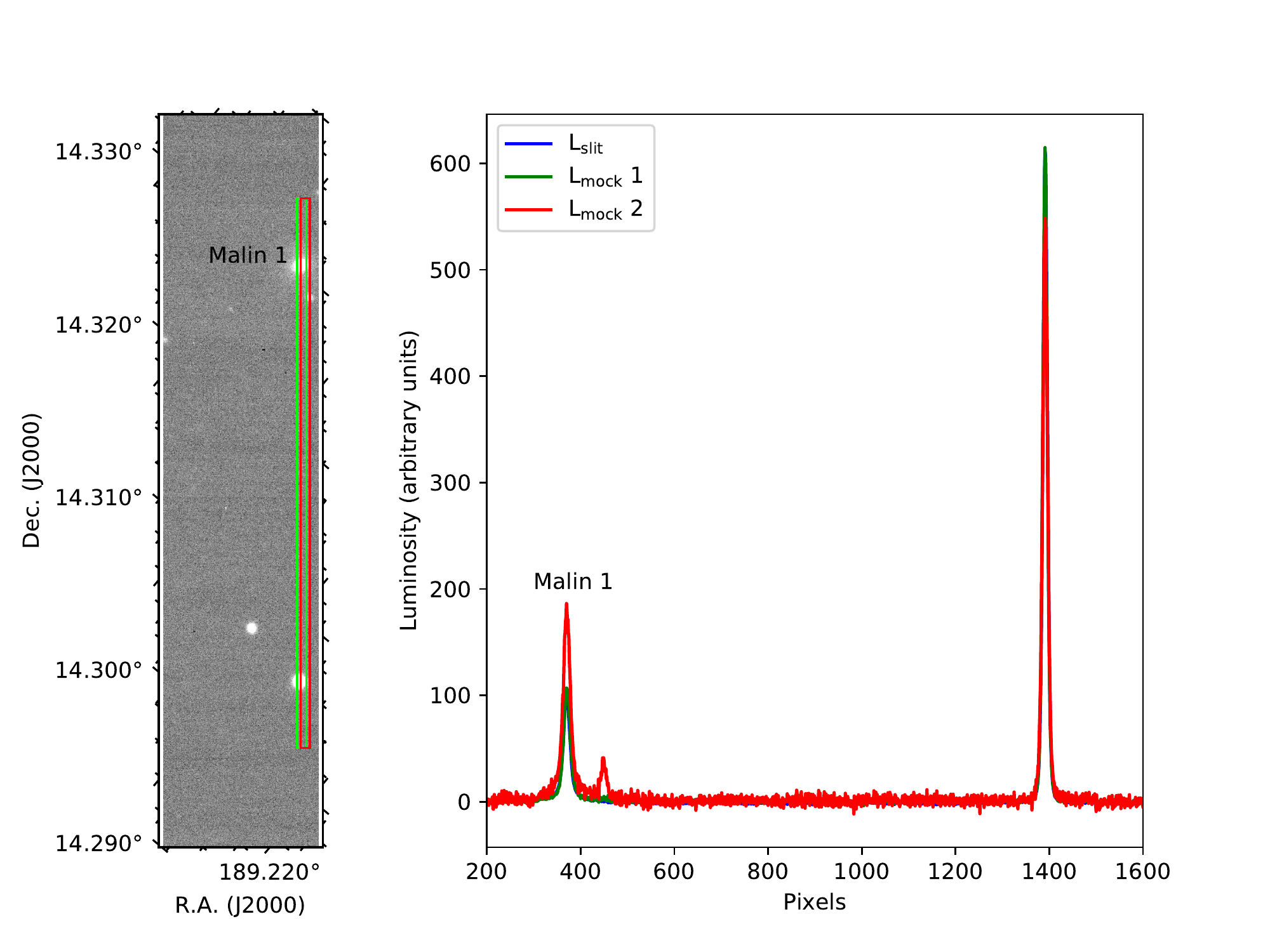}
    \caption{\textit{Left:} Two mock slits for the slit 1 of our 2016 observations (see Fig. \ref{slits}), placed on the image of the galaxy acquired during the night of the observation. The two mock slits (green and red) are shifted apart by 1$\arcsec$, with the green slit marking the best slit position obtained in our simulation. \textit{Right:} The luminosity profile along the two mock slits shown in the left panel (from top to bottom). The blue curve is the luminosity distribution passing through the slit obtained from the 2D spectrum ($L_{\mathrm{slit}}(pixel)$) of the slit 1 observation (integrated over wavelength to be consistent with the image).}
    \label{mock_slits}
    \end{figure}

    An additional source of error in our measurements comes from the uncertainty in the precise positioning of the slit on the sky during each observation. As discussed in Sect. \ref{data_and_reduction}, we had extracted spectra from three different slit positions for the \Ha{} data and a single slit position for the \oii{} data. In order to obtain a precise slit position for each observations and the associated uncertainties, we simulated the expected luminosity along the slit on the basis of an image of the galaxy acquired during the 2016 observation, just before the spectroscopic observations (see Fig. \ref{mock_slits}). 
    The 2D spectrum at each slit position was collapsed along its spectral axis in order to obtain the full luminosity distribution passing through the slit ($L_{\mathrm{slit}}(pixel)$). 
    A mock slit luminosity distribution $L_{\mathrm{mock}}(pixel,x,y)$ is computed from the image for any slit position $(x,y)$, using the same slit width and angle as used in the observations . 
    We started from the position of the slit expected at the telescope during our observations, and explored shifts in (x,y) around this position, in order to find the best pixel position for the slit.

    \begin{equation}\label{slit_position_chisquare}
        \chi^{2}(x,y)=\sum_{slit \, pixels}^{} \frac{\Bigg\{L_{\mathrm{slit}}(pixel) - \left[C \times L_{\mathrm{mock}}(pixel,x,y)\right]\Bigg\}^2}{\sigma^{2}_{\mathrm{pixel}}}
    \end{equation}

    where $\sigma^{2}_{\mathrm{pixel}}$ is the total sky level noise and pixel scale Poisson noise measured from $L_{\mathrm{slit}}(pixel)$. 
    The coefficient $C$ is the matching coefficient, computed from the $L_{\mathrm{slit}}$ and $L_{\mathrm{mock}}$ by the following relation:
    
    \begin{equation}
        C = \Bigg| \frac{\sum_{slit \, pixels}^{} \left[L_{\mathrm{slit}}(pixel) \times L_{\mathrm{mock}}(pixel,x,y)\right]}{\sum_{slit \, pixels}^{} L^{2}_{\mathrm{mock}}(pixel,x,y)} \Bigg|
    \end{equation}
        
    Upon minimizing Eqn. \ref{slit_position_chisquare}, we obtained a pixel position (x,y) for each of the slit position with a minimum value of $\chi^{2}_{\mathrm{min}}$. 
    The upper and lower confidence levels in the pixel positions were obtained by incrementing the $\chi^{2}_{\mathrm{min}}$ value with a $\Delta\chi^{2}$ value of 6.63 to get a $99\%$ confidence level (see Table. 1 of \citealt{avni}). 
    For slit positions 1, 2 and 3 of the 2016 observations, this slit positioning uncertainties were found to be 3 pixels, 1 pixel and 1 pixel respectively. For the slit position of the 2019 observation, since we do not have an image on the same night to perfrom the same calculation, we used the maximum uncertainty of 3 pixels from the 2016 observations to account for this uncertainty (considering the observational setup is the same, we think this will cover the positioning error of the telescope and spectrograph).
    
    De-projecting the above mentioned pixel uncertainties on the sky provides a maximum error of $\sim0.3\arcsec$, and typically of order $0.1 \arcsec$. This uncertainty in the sky coordinates of our regions were propagated while measuring the radius, azimuth angle ($\cos{\theta}$) and V$_\mathrm{rot}$ on the galaxy plane (radius and $\cos{\theta}$ were computed from the sky coordinates using \textit{astropy.coordinates.SkyCoord} python package).

\subsection{Combining the error}
   
    Since V$_\mathrm{rot}$ on the galaxy plane depends on both the observed wavelength $\lambda_{\mathrm{obs}}$ and the de-projection angle $\cos{\theta}$, it is affected by both the wavelength fitting error and the error in $\cos{\theta}$ due to the slit positioning uncertainty. We made a quadratic sum of both of theses error contributions to obtain our final error in V$_\mathrm{rot}$ shown in this work. In most cases, the error is in fact dominated by the positioning error.

    Obviously, the wavelength and flux uncertainties are only affected by the fitting error, whereas the radius values is only affected by the slit positioning error (see Table. \ref{data}).

   \section{Malin 1 Nuclear activity \& Metallicity indication}
    \label{BPT}         
\begin{figure}[h]
        \centering
        \includegraphics[width=\hsize]{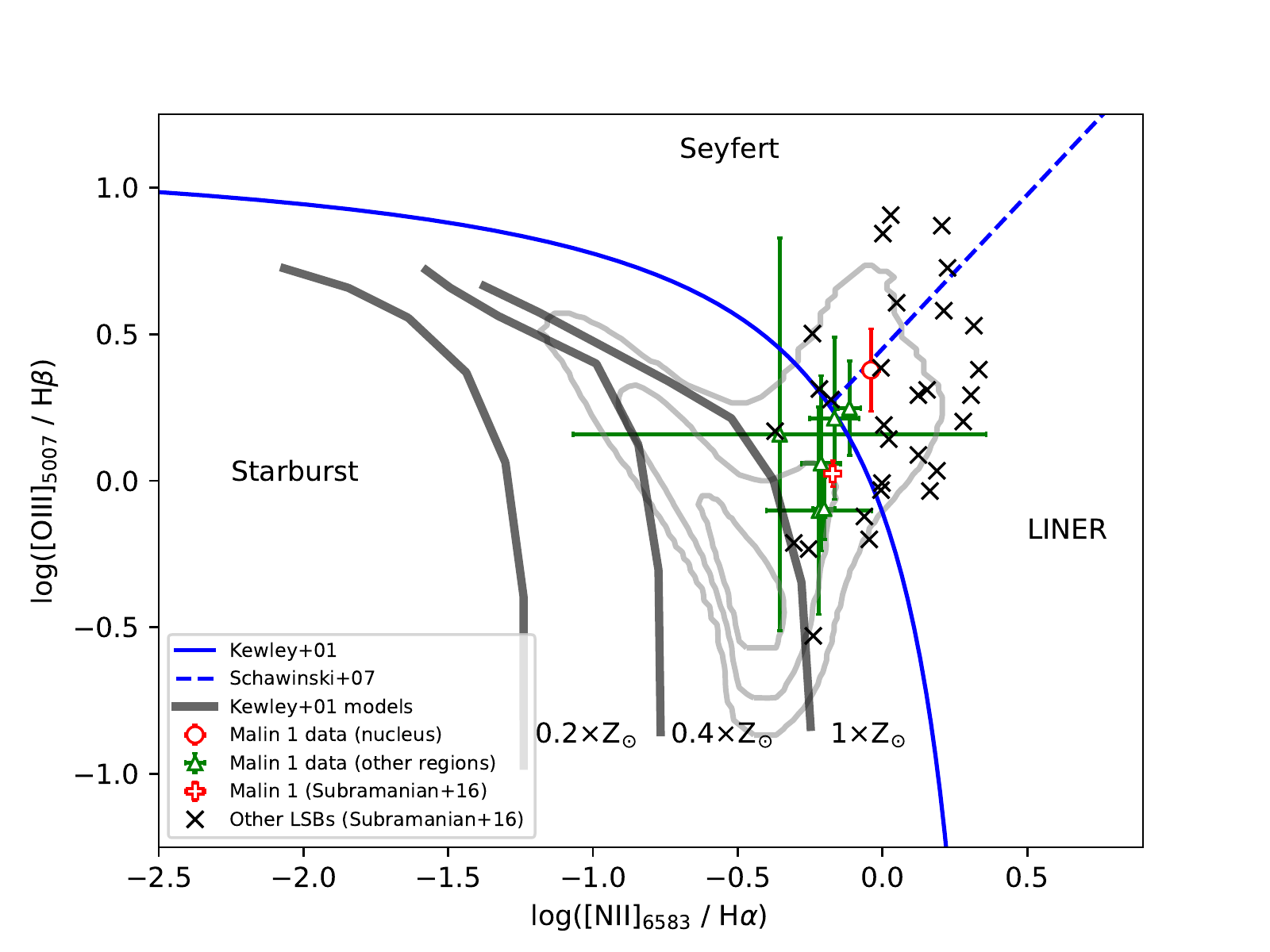}
        \caption{
        Line diagnostic BPT diagram with $\log$(\niib{}/\Ha{}) vs. $\log$([\ion{O}{iii}]$_{5007}$/\Hb{}). The red circle and the green triangles indicate the data points of Malin 1 respectively for the nucleus and other detected regions in this work. The red plus and black crosses are the LSB sample from \citet{subramanian16}, including Malin 1. The blue solid and dashed lines are defined by \citet{Kewley2001} and \citet{Schawinski2007} respectively for the separation of AGN from star forming regions. The gray contours show the distribution of a random sample of nuclear spectra of SDSS galaxies in the redshift range 0.01–1 and stellar mass $10^{9}$-$10^{11}$ \citep{vestige_bpt}. The black solid thick lines shows the expected behavior of star forming regions as derived from the photo-ionization models of \citet{Kewley2001} for three different metallicities (0.2, 0.4, 1 $Z_{\odot}$).
        }
        \label{BPT_fig}
        \end{figure}

    \cite{barth} classifies Malin 1 as a LINER nucleus galaxy with an \niib{}/\Ha{} flux ratio of 0.85. This is close to our measured flux ratio of \niib{}/\Ha{} = $0.91\pm0.06$  in the central region of Malin 1. \cite{subramanian16} gives a similar classification for Malin 1, placing it in the category of a LINER and composite nuclei with weak ionization contributions from both AGN and starbursts on the basis of several diagnostics. Our measured central \niib{}/\Ha{} and [\ion{O}{iii}]$_{5007}$/\Hb{} flux ratios would place Malin 1 in the borderline of LINER-Seyfert classification in a BPT diagram \citep{bpt, Kewley2006} as shown in Fig. \ref{BPT_fig}. 
    Other LSBs from \citet{subramanian16} are located in a similar place as Malin 1 in this diagram.
    The flux ratios from the other detected regions (6 regions in addition to the nucleus) lies in the starburst region but close to the starburst-AGN demarcation line.
    We also show in Fig. \ref{BPT_fig} photo-ionisation models of \citet{Kewley2001} showing that the inner regions of Malin 1 may have a large metallicity. Using the calibration of \citet{pettini}, the corresponding \niib{}/\Ha{} ratio indeed point to metallicities about 0.15 dex above solar \citep[or almost solar if we use the improved calibration of][]{Marino2013}. Such metallicities are close to the values expected from the stellar-mass metallicity relationship derived by \cite{Bian2016} for the SDSS galaxies using the same metallicity indicator (and adopting $log(M_*/M_{\odot})=10.87$ for Malin 1 after integrating the profile presented in Sect. \ref{mass_model}). On the other hand, the high metallicity is at odd with the large amount of gas, but low attenuation (discussed in Sect. \ref{subsecSFRmeasurement}), and absence of detection in CO \citep{Braine2000}.

\end{appendix}

\end{document}